\documentclass[a4paper,16pt]{article}
\usepackage[utf8]{inputenc}
\usepackage[english]{babel}
\usepackage{graphicx}
\usepackage{mathtext}
\usepackage{indentfirst}
\usepackage{epsfig,amsmath,amsfonts,braket,fp,bbm,tikz-cd,authblk,hyperref}
\usepackage{amssymb}
\usepackage{slashed}
\usepackage{euscript}
\usepackage{bbold}
\usepackage{bm,paralist,xspace,url,relsize}
\usepackage{color,xcolor}
\usepackage{hyperref}
\usepackage{textcomp}
\usepackage{titlesec}
\usepackage{blindtext}
\usepackage{cmap} 
\usepackage[left=2cm,right=2cm,top=2cm,bottom=2cm,bindingoffset=0cm]{geometry}
\PassOptionsToPackage{unicode}{hyperref}
\PassOptionsToPackage{naturalnames}{hyperref}
\usepackage{titlesec}
\titleformat{\section}{\normalfont\large\bfseries}{\thesection}{1em}{}
\titleformat{\subsection}{\normalfont\normalsize\bfseries}{\thesubsection}{1em}{}

\usepackage[active]{srcltx}

\usepackage{tikz}

\begin{document}
\title{\textbf{Massive scalar field theory in the presence of moving mirrors}}

\renewcommand\Affilfont{\itshape\footnotesize}
\renewcommand\Authands{, }
\setcounter{footnote}{3}

\author[1,2]{L. Astrakhantsev}
\author[3]{O. Diatlyk}

\affil[1]{B. Cheremushkinskaya, 25, Institute for Theoretical and Experimental Physics, 117218, Moscow, Russia}
\affil[2]{Institutskii per, 9, Moscow Institute of Physics and Technology, 141700, Dolgoprudny, Russia}
\affil[3]{National Research University Higher School of Economics, Department of
Mathematics and International Laboratory of Representation Theory and Mathematical
 Physics, Russia}

\date{}
\maketitle

\begin{abstract}
\noindent
We study the 2D massive fields in the presence of moving mirrors. We do that for standing mirror and mirror moving with constant velocity. We calculate the modes and commutation relations of the field operator with the corresponding conjugate momentum in each case. We find that in case of the ideal mirror, which reflects modes with all momenta equally well, the commutation relations do not have their canonical form. However, in the case of non--ideal mirror, which is transparent for the modes with high enough momenta, the commutation relations of the field operator and its conjugate momentum have their canonical form. Then we calculate the free Hamiltonian and the expectation value of the stress--energy tensor in all the listed situations. In the presence of moving mirrors the diagonal form in terms of the creation and annihilation operators has the operator that performs translations along the mirror's world line rather than the one which does translations along the time--line. For the massive fields in the presence of a mirror moving with constant velocity the expectation value of the stress--energy tensor has a non--diagonal contribution which decays with the distance from the mirror.

\end{abstract}

 \setcounter{section}{0}
\renewcommand{\contentsname}{Contents}
\begin{titlepage}

	\end{titlepage}

\newpage

\tableofcontents

\newpage
\section{Introduction}

Quantum field theory in non--stationary situation recently received a lot of attention. The main interest in the context of particle physics and cosmology is in the particle creation and the resulting back reaction. The common wisdom is that semi--classical approximation provides the leading result for the flux. In the semi--classical approximation the flux, if any, appears due to the amplification of the zero--point fluctuations.

However, in many situations there is a break down of the perturbation theory and of the semi--classical approximation. Namely there is a universal secular growth of the loop corrections to the correlation functions, which appears due to the excitation of the higher levels on top of the above mentioned amplification of the zero point fluctuations.

Such a secular growth is observed in de Sitter space quantum field theory \cite{AkhmedovKEQ}--\cite{Akhmedov:2017ooy}, in strong electric fields \cite{Akhmedov:2014hfa}, \cite{Akhmedov:2014doa}, in the presence of the black hole collapse \cite{AkhGodPop} and in the presence of the moving mirrors \cite{Akhmedov:2017hbj}. The presence of the secular effects, which make quantum corrections strong, demands the resummation of the leading contributions from all loops. The situation with mirrors seems to be the simplest one among the listed in the previous paragraph phenomena.

However, the authors of \cite{Akhmedov:2017hbj} have encountered some unexpected problems. One of them is related to the infrared peculiarities of the massless two--dimensional fields. The second problem appears due to the consideration of the ideal mirror \cite{Akhmedov:2018lkp} --- of such a mirror which reflects all modes equally well, independently of the value of their momenta. Meanwhile a real physical mirror is certainly transparent for high enough momenta modes.

As explained in \cite{Akhmedov:2018lkp}, due to the reflection of all momenta the character of the UV divergences in non--globaly hyperbolic situations in the Minkowskian signature is very much different from the one in empty space or in Euclidian signature.

To overcome the above difficulties with massless fields in this paper we consider massive 2D scalar quantum field theory in the presence of mirrors performing various types of motion. The action for the massive scalar field that we consider is as follows:

$$S=\int\limits_{x\geqslant z(t)} dtdx\Big[\eta^{\mu\nu}\partial_{\mu}\phi\partial_{\nu}\phi-m^2\phi^2\Big],$$ where $\eta_{\mu\nu}=diag(1,-1)$ is the metric for two-dimensional Minkowski space-time.

In this paper we consider free fields. The situation with loop corrections in the $\phi^4$ theory, along the lines of \cite{Akhmedov:2017hbj} will be considered elsewhere. Also we consider here only the case of either standing mirrors or moving with constant velocity. The case of mirrors performing non--inertial motions will also be considered elsewhere.

Ideal mirror in this theory introduced as the following boundary condition:
\begin{equation}
\phi(t,z(t))=0,
\end{equation}
which is the time-like curve --- the mirror's world--line $z(t)$.

To overcome the aforementioned difficulties with non--ideal mirrors we also consider non-ideal mirror. The action in this case is
$$S=\int dtdx\Big[\eta^{\mu\nu}\partial_{\mu}\phi\partial_{\nu}\phi-m^2\phi^2-V(x)\phi^2\Big],$$
and the equations of motion are:
\begin{equation}
\Big(\partial^2_{t}-\partial^2_{x}+m^2+V(x)\Big)\phi(t,x)=0.
\end{equation}
The potential barrier $V(x)$ plays the role of the mirror. In particular we will consider $V(x)=\alpha\delta(x)$. As is explained in the paper such a mirror is transparent for high enough energy modes and reflects low energy ones. The scale is set by the parameter $\alpha$. Similar situation for the massless fields has been considered in \cite{Bordag:1992cm}, \cite{Munoz-Castaneda:2013yga}.

The paper is organized as follows. In the section 2 we consider the situation with the massless fields to set up the notations. Free 2D massless fields in the presence of moving mirrors have been considered in many places (see e.g. \cite{1}, \cite{3}).

First, we find the modes of the massless field in the presence of standing ideal mirror. Second, we check the commutation relations of the field operator and its conjugate momentum and find that they are different from the canonical ones. This is true under the assumption that the creation and annihilation operators obey the standard Heisenberg algebra.  These problems appear due to the fact that the mirror is ideal. Third, we derive the free Hamiltonian via the creation and annihilation operators and the expectation value of the stress--energy tensor. Then, in the section 2 we continue with the consideration of the mirror moving with constant velocity and perform the same steps as for the standing mirror. We point out the correct form of the free Hamiltonian which is different from its usual form.

In the section 3 we continue with the consideration of the massive fields in the presence of moving mirrors.
Here we also calculate the commutation relations of the field operator with the corresponding conjugate momentum. We do that for standing mirror and mirror moving with constant velocity and find that they do not have the canonical form. Then we calculate the free Hamiltonian and the expectation value of the stress--energy tensor.

In the section 4 we perform the same steps for the massive fields in the presence of standing and moving non--ideal mirrors. In the case of non--ideal mirror we find that the field operator and its conjugate momentum obey the canonical commutation relations. We conclude in the section 5.

\section{Massless scalar field in presence of an ideal mirror}

To set up the notations we start with the consideration of the massless scalars in the presence of a perfect mirror. This is such a mirror that reflects modes with all momenta equally well.
\subsection{Mirror at rest}
Consider the case of massless scalar field with Dirichlet boundary condition $\phi(t,x)=0$. Equation of motion for this field is
\begin{equation}
\partial^2_{t}\phi(t,x)-\partial^2_{x}\phi(t,x)=0.
\end{equation}
In the presence of mirror the mode expansion is as follows:
\begin{equation}
\label{em}
\phi(t,x)=iA\int\limits_{0}^{+\infty} \frac{dk}{\sqrt{2k}}\Big[a_{k}\sin(kx)e^{-ikt}-a_{k}^\dagger \sin(kx)e^{ikt}\Big].
\end{equation}
If one uses the standard Heisenberg algebra for $a_{k}$ and $a_{k}^\dagger$ and sets $A=\frac{1}{\pi}$, then for the field $\phi(t,x)$ and its conjugate momentum $\pi(t,y)=\partial_{t}\phi(t,y)$ one finds the following commutation relations:
\begin{equation}
[\phi(x),\pi(y)]=i\Big\{\delta(x-y)-\delta(x+y)\Big\},
\end{equation}
where in addition to the usual delta-function appears the boundary one $\delta(x+y)$, whose argument is equal to zero only on the mirror surface ($x=y=0$). The presence of such a boundary delta-function is related to the absence of global hyperbolicity due to the mirror. Note that the quantum field theory in question is defined for $x,y\geqslant0$. First time such an event was encountered in \cite{Akhmedov:2017hbj}. We will discuss this point in the concluding section.

The Hamiltonian of the system is as follows:
$$H=\int\limits_{0}^{+\infty}T_{tt}dx=\frac{1}{2}\int\limits_{0}^{+\infty}\Big[\pi^2+(\partial_{x}\phi)^2\Big]dx,$$
or in terms of $a_{k}$ and $a_{k}^\dagger$:
\begin{equation}
H=\int\limits_{0}^{+\infty}\frac{dk}{2\pi}\frac{k}{2}\Big(a_{k}a_{k}^\dagger+a_{k}^\dagger a_{k}\Big).
\end{equation}
This Hamiltonian has the ordinary diagonal form. The only difference with respect to the empty space case is that now $k\geq0$ (see also eq. $\eqref{em}$).

In this paper we are interested in the calculation of the expectation value of the stress-energy tensor:
\begin{equation}
\label{tmn}
T_{\mu\nu}=\frac{1}{2}\{\partial_{\mu}\phi,  \partial_{\nu}\phi\}-\frac{1}{2}\eta_{\mu\nu}\Big[\frac{1}{2}\eta^{\alpha\beta}\{\partial_{\alpha}\phi,  \partial_{\beta}\phi\}-m^2\phi^2\Big],
\end{equation}
where $\{$ $,\}$ is the anticommutator. E.g. the vacuum expectation value of the flux is
\begin{equation}
\braket{T_{tx}}=\frac{1}{2}\Big(\braket{\partial_{x}\phi(x,t),\pi(x,t)}+\braket{\pi(x,t),\partial_{x}\phi(x,t)}\Big),
\end{equation}
which is of particular interest for us and in the case under consideration it is equal to
\begin{equation}
\braket{T_{tx}}=\frac{1}{2}\int\limits_{0}^{+\infty}\frac{dk}{2\pi}\int\limits_{0}^{+\infty}\frac{dk'}{2\pi}\frac{(\sin{k'x}\cos{kx}-\sin{kx}\cos{k'x})}{2\sqrt{kk'}}\delta(k-k')=0.
\end{equation}
Below we will encounter situations when the flux is not zero.
\subsection{Mirror moving with constant velocity}
In the presence of the moving mirror from the boundary condition $\phi(t,-\beta t)=0$ and the Klein-Gordon equation $\partial^2_{t}\phi(t,x)-\partial^2_{x}\phi(t,x)=0$ one can find the mode functions and write down the field operator via the light cone coordinates $u=t-x$ and $v=t+x$ as follows:
\begin{equation}
\phi(t,x)=i\int\limits_{0}^{+\infty} \frac{dk}{2\pi}\frac{1}{\sqrt{2k}}\Bigl[a_{k}(e^{-ikv}-e^{-ik\Omega u})-a_{k}^\dagger(e^{ikv}-e^{ik\Omega u})\Bigr], \quad \Omega=\frac{1-\beta}{1+\beta}.
\end{equation}
The commutation relations of the field $\phi(t,x)$ and its conjugate momentum $\pi(t,y)$ are:
 \begin{equation}
 [\phi(x),\pi(y)]=i\Big\{\delta(x-y)-\frac{\Omega}{2}\delta\Big[x+\Omega y+t(1-\Omega)\Big]-\frac{1}{2}\delta\Big[t(\Omega-1)-\Omega x- y\Big]\Big\}.
 \end{equation}
The last two delta-functions on the RHS are the boundary ones, i.e. relevant only for $x=y=-\beta t$.

The Hamiltonian is defined as:
\begin{equation}
H=\frac{1}{2}\int\limits_{-\beta t}^{+\infty}dx\Big[\pi^2+(\partial_{x}\phi)^2\Big].
\end{equation}
Direct calculation shows that it has non-diagonal $a_{k}a_{k}$ and $a_{k}^{\dag}a_{k}^{\dag}$ contributions. Moreover, the coefficients of these non-diagonal parts are time-dependent. Namely, in terms of $\hat{a}_{k}$ and $\hat{a}_{k}^{\dagger}$ operators the Hamiltonian contains the following terms:
$$\Delta H=\frac{i\beta}{1+\beta}\int\limits_{0}^{+\infty}\int\limits_{0}^{+\infty}\frac{dk}{2\pi}\frac{dk'}{2\pi}\frac{\sqrt{kk'}(k+k')e^{-i(k+k')(1-\beta)t}}{(k+k')^2+\epsilon^2}a_{k}a_{k'}+h.c.,$$
where we have introduced the regularization parameter $\epsilon$ to regulate the integration over $x$.

The same sort of non-diagonal contributions has also the momentum operator of the system $P=\int\limits_{-\beta t}^{+\infty}dxT_{tx}$. However, the operator which is responsible for the translations along the mirror world-line, i.e.
\begin{equation}
\label{HH}
H_{proper}=H-\beta P,
\end{equation}
is diagonal and time independent.
In fact, in terms of $\hat{a}_{k}$ and $\hat{a}_{k}^\dagger$ the proper Hamiltonian has the following form:
\begin{equation}
H_{proper}=(1-\beta)\int\limits_{0}^{+\infty}\frac{dk}{2\pi}\frac{k}{2}\Big(a_{k}a^\dagger_{k}+a^\dagger_{k}a_{k}\Big).
\end{equation}
Such a discussion will be important for the case of massive fields. Furthermore, to compute the vacuum expectation value of the flux, we use the point-splitting regularization. The result is
\begin{equation}
\braket{T_{tx}}=\lim_{\epsilon\to 0}\  \frac{1}{2}\Big<\partial_t\phi(t,x)\partial_x\phi(t+i\epsilon,x)+\partial_x\phi(t,x)\partial_t\phi(t+i\epsilon,x)\Big>=\lim_{\epsilon\to 0}\int\limits_{0}^{+\infty}\frac{dk}{2\pi}\frac{k}{2}\Big(e^{-k\epsilon}-\Omega^2 e^{-k\Omega\epsilon}\Big)=0.
\end{equation}
As it should be, mirror moving with constant velocity does not generate a particle flux.

\section{Massive scalar field in the presence of an ideal mirror}

We continue our considerations with the discussion of massive fields in the presence of standing and moving ideal mirrors.

\subsection{Mirror at rest}

First, consider the case of the simplest boundary condition, i.e. the case of mirror at rest. In such a case the modes are as follows:
\begin{equation}
\phi(t,x)=i\int\limits_{0}^{+\infty}\frac{dk}{2\pi}\sqrt{\frac{2}{\omega}}\Big(a_{k} \sin(kx)e^{-i\omega t}-a_{k}^\dagger \sin(kx)e^{i\omega t}\Big),
\end{equation}
and the conjugate momentum is
\begin{equation}
\pi(t,x)=\partial_{t}\phi(t,x)=\int\limits_{0}^{+\infty}\frac{dk}{2\pi}\sqrt{2\omega}\Big(a_{k} \sin(kx)e^{-i\omega t}+a_{k}^\dagger \sin(kx)e^{i\omega t}\Big),
\end{equation}
where the dispersion relation for the massive field is $\omega=\sqrt{k^2+m^2}$.

The commutation relations between $\phi(t,x)$ and $\pi(t,y)$ are as follows:
\begin{equation}
[\phi(t,x),\pi(t,y)]=i\Big\{\delta(x-y)-\delta(x+y)\Big\},
\end{equation}
if we assume the standard Heisenberg algebra for $a_{k}$ and $a_{k}^\dagger$. Thus, similarly to the case of the massless field on top of the ordinary delta-function there is the boundary one.

The Hamiltonian has the ordinary diagonal form:
\begin{equation}
H=\frac{1}{2}\int\limits_{0}^{+\infty}\Big(\pi^2+(\partial_{x}\phi)^2+m^2\phi^2\Big)dx=\int\limits_{0}^{+\infty}\frac{dk}{2\pi}\frac{\omega}{2}\Big(a_{k}a_{k}^\dagger+a_{k}^\dagger a_{k}\Big).
\end{equation}
Please note that the region of integration over the momenta is $k\geqslant0$, unlike the case of empty space, where $k\in(-\infty,+\infty)$.

The vacuum expectation value of the flux is
\begin{equation}
\braket{T_{tx}}=\frac{1}{2}\int\limits_{0}^{+\infty}\frac{dk}{2\pi}\int\limits_{0}^{+\infty}\frac{dk'}{2\pi}\frac{(\sin{k'x}\cos{kx}-\sin{kx}\cos{k'x})}{2\sqrt{\omega\omega'}}\delta(k-k')=0,
\end{equation}
as it should be in the presence of standing mirror.

Now let us turn to the less trivial case of mirror moving with constant velocity.

\subsection{Mirror moving with constant velocity}

\subsubsection{Modes}

Consider a mirror which moves with a velocity $\beta<1$. To find the modes in such a case we do the Lorentz boost $(x,t)\to(x',t')$.
After the boost the modes change as
\begin{equation}
h(t,x)=iA\big(e^{-i\omega_{+} t'-ik_{+}x'}-e^{-i\omega_{-} t'+ik_{-}x'}\big),
\end{equation}
where $\omega_{+}=\omega\gamma+\beta\gamma k$, $\omega_{-}=\omega\gamma-\beta\gamma k$, $k_{+}=\gamma k+\beta\gamma\omega$ and $k_{-}=\gamma k-\beta\gamma\omega$. The minimal value of $\omega_{+}$ and $k_{+}$ is $\gamma m$ and $\beta\gamma m$, correspondingly, $(\omega_{+},k_{+})$ and $(\omega_{-},k_{-})$ are the energy and momentum of the waves that are falling and reflected from the mirror, correspondingly.

In the following we change the notations as $(\omega_{+},k_{+})\to(\omega,k)$, $(\omega_{-},k_{-})\to(\omega_{r},k_{r})$.
The frequency and the wave vector of the falling and reflected waves obey the following relation:

\begin{equation}
\omega_{r}=(1+\beta^2)\gamma^2\omega-2\beta\gamma^2k,
\end{equation}
\begin{equation}
k_{r}=-2\beta\gamma^2\omega+(1+\beta^2)\gamma^2k,
\end{equation}
where the reflected wave obviously has the same dispersion relation $k_{r}^2+m^2=\omega_{r}^2$ and $\gamma^2=\frac{1}{1-\beta^2}$.

The field operator is
\begin{equation}
\label{k}
\phi(t,x)=iA\int\limits_{\gamma\beta m}^{+\infty} \frac{dk}{2\pi}\frac{1}{\sqrt{2\omega}}\Big[a_{k}\big(e^{-i\omega t-ikx}-e^{-i\omega_{r} t+ik_{r}x}\big)-a_{k}^\dagger\big(e^{i\omega t+ikx}-e^{i\omega_{r} t-ik_{r}x}\big)\Big],
\end{equation}
where $A$ is a normalization constant, to be determined from the commutation relations. It is straightforward to check that modes obey the K-G equation and the appropriate boundary conditions.

Note that the integration over the momentum $k$ in the integral ($\ref{k}$) is cut from below by $\gamma\beta m$, which physically corresponds to the fact that modes with momenta $k<\gamma\beta m$ cannot catch up with the mirror, i.e. for lower values of momenta the reflected wave simply does not exist.

Also, it is easy to see that when $\omega=k, m=0$ the modes in $\eqref{k}$ reduce to their massless form, which was considered above.

\subsubsection{Commutation relations}

In this section we check if the field and its conjugate momentum obey the canonical commutation relations or not. The conjugate momentum to the field $\eqref{k}$ is
\begin{equation}
\pi(t,x)=A\int\limits_{\gamma\beta m}^{+\infty} \frac{dk}{2\pi}\frac{1}{\sqrt{2\omega}}a_{k}\Big[\omega e^{-i\omega t-ikx}-\omega_{r}e^{-i\omega_{r}t+ik_{r}x}\Big]+h.c.
\end{equation}
Then the commutation relations are as follows:
$$[\phi(t,x), \pi(t,y)]=iA^2\int\limits_{\gamma\beta m}^{+\infty} \frac{dk}{2\pi}\frac{1}{2\omega}\Big(e^{-i\omega t-ikx}-e^{-i\omega_{r} t+ik_{r}x}\Big)\Big[\omega e^{i\omega t+iky}-\omega_{r}e^{i\omega_{r}t-ik_{r}y}\Big]+iA^2\cdot h.c.=$$$$=
\frac{iA^2}{2}\int\limits_{\gamma\beta m}^{+\infty} \frac{dk}{2\pi}\Big(e^{ik(y-x)}+\frac{\omega_{r}}{\omega}e^{ik_{r}(x-y)}\Big)-
\frac{iA^2}{2}\int\limits_{\gamma\beta m}^{+\infty}\frac{dk}{2\pi}\Big(e^{-i(\omega_{r}-\omega)t+ik_{r}x+iky}+\frac{\omega_{r}}{\omega}e^{i(\omega_{r}-\omega)t-ikx-ik_{r}y}\Big)+\frac{iA^2}{2}\cdot h.c.$$
It can be checked that $\frac{dk}{\omega}=\frac{dk_{r}}{\omega_{r}}$, using this we find that
$$\int\limits_{\gamma\beta m}^{+\infty} \frac{dk}{2\pi}\Big(e^{ik(y-x)}+\frac{\omega_{r}}{\omega}e^{ik_{r}(x-y)}\Big)=\int\limits_{\gamma\beta m}^{+\infty} \frac{dk}{2\pi}e^{ik(y-x)}+\int\limits_{-\gamma\beta m}^{+\infty} \frac{dk_{r}}{2\pi}e^{ik_{r}(x-y)}= \\ \\\int\limits_{-\infty}^{+\infty} \frac{dk}{2\pi}e^{ik(x-y)}=\delta(x-y).$$
Then if $A=1$, the commutation relations have the form
$$[\phi(x),\pi(y)]=i\delta(x-y)-\frac{i}{2}\int\limits_{\gamma\beta m}^{+\infty}\frac{dk}{2\pi}\Big[e^{-i(\omega_{r}-\omega)t+ik_{r}x+iky}+\frac{\omega_{r}}{\omega}e^{i(\omega_{r}-\omega)t-ikx-ik_{r}y}\Big]-$$$$-\frac{i}{2}\int\limits_{\gamma\beta m}^{+\infty}\frac{dk}{2\pi}\Big[e^{i(\omega_{r}-\omega)t-ik_{r}x-iky}+\frac{\omega_{r}}{\omega}e^{-i(\omega_{r}-\omega)t+ikx+ik_{r}y}\Big].$$
The last integral can be transformed into the following form
$$\int\limits_{\gamma\beta m}^{+\infty}\frac{dk}{2\pi}\Big(e^{-i(\omega_{r}-\omega)t+ik_{r}x+iky}+\frac{\omega_{r}}{\omega}e^{i(\omega_{r}-\omega)t-ikx-ik_{r}y}\Big)=\int\limits_{\gamma\beta m}^{+\infty}\frac{dk}{2\pi}e^{-i(\omega_{r}-\omega)t+ik_{r}x+iky}+$$$$+\int\limits_{-\gamma\beta m}^{+\infty}\frac{dk_{r}}{2\pi}e^{i(\omega_{r}-\omega)t-ikx-ik_{r}y}=\big\{k_{r}\to-k_{r}, k\to -k\big\}=\int\limits_{-\infty}^{+\infty}\frac{dk}{2\pi}e^{-i(\omega_{r}-\omega)t+ik_{r}x+iky},$$
where to obtain the last equality we have used that $k_{r}$ is a homogeneous function of $k$.

Hence,
\begin{equation}
\label{cr}
[\phi(x),\pi(y)]=i\delta(x-y)-\frac{i}{2}\int\limits_{-\infty}^{+\infty}\frac{dk}{2\pi}e^{-i(\omega_{r}t-k_{r}x)+i(\omega t+ky)}-\frac{i}{2}\int\limits_{-\infty}^{+\infty}\frac{dk}{2\pi}e^{i(\omega_{r}t-k_{r}x)-i(\omega t+ky)}.
\end{equation}
If we introduce the notations $A=2\gamma^2\beta t+(1+\beta^2)\gamma^2 x+y$ and $B=2\gamma^2\beta^2t+2\beta\gamma^2x$ and use the regularization $B\to B-i\epsilon$, as $\epsilon\to 0$, then
$$\int\limits_{-\infty}^{+\infty}dke^{iAk-(iB+\epsilon)\omega}dk=2\int\limits_{0}^{+\infty}dk\cos(Ak)e^{-(iB+\epsilon)\sqrt{k^2+m^2}}.$$
Similarly one can represent the second integral on the RHS of ($\ref{cr}$) as:
$$\int\limits_{-\infty}^{+\infty}\frac{dk}{2\pi}e^{i(\omega_{r}t-k_{r}x)-i(\omega t+ky)}=2\int\limits_{0}^{+\infty}dk\cos(Ak)e^{-(-iB+\epsilon)\sqrt{k^2+m^2}}.$$
Using the following table integral:
\begin{equation}
\int\limits_{0}^{+\infty}dk\cos(Ak)e^{-C\sqrt{k^2+m^2}}=\frac{Cm}{\sqrt{A^2+C^2}}K_{1}(m\sqrt{A^2+C^2})\quad{\rm where}\quad Re(C),Re(m)\textgreater0,
\end{equation}
where $K_{1}(x)$ is the modified Bessel function of the second kind, we obtain that the two integrals on the RHS of ($\ref{cr}$) are equal to
$$2m\epsilon\Big[\frac{K_{1}(m\sqrt{A^2+(iB+\epsilon)^2})}{\sqrt{A^2+(iB+\epsilon)^2}}+\frac{K_{1}(m\sqrt{A^2+(\epsilon-iB)^2})}{\sqrt{A^2+(\epsilon-iB)^2}}\Big]+$$$$+2iBm\Big[\frac{K_{1}(m\sqrt{A^2+(iB+\epsilon)^2})}{\sqrt{A^2+(iB+\epsilon)^2}}-\frac{K_{1}(m\sqrt{A^2+(\epsilon-iB)^2})}{\sqrt{A^2+(\epsilon-iB)^2}}\Big].$$
To evaluate this expression, first, consider the case of $x,y\neq -\beta t$; $A,B\neq 0$.
Then the first term in the sum is zero because it is just a product of $\epsilon\to 0$ and some finite number.
At the same time the second term is
$$2iBm\Big[\frac{K_{1}(m\sqrt{|A^2-B^2+\epsilon^2+2iB\epsilon|}e^{i\frac{1}{2}(\varphi+2\pi n)})}{\sqrt{|A^2-B^2+\epsilon^2+2iB\epsilon|}e^{i\frac{1}{2}(\varphi+2\pi n)}}-\frac{K_{1}(m\sqrt{|A^2-B^2+\epsilon^2-2iB\epsilon|}e^{i\frac{1}{2}(-\varphi+2\pi n)})}{\sqrt{|A^2-B^2+\epsilon^2-2iB\epsilon|}e^{i\frac{1}{2}(-\varphi+2\pi n)}}\Big],$$
where the square root is understood as the multi-valued function and $\varphi=\arctan{\big[\frac{2\epsilon B}{A^2-B^2+\epsilon^2}\big]}$. Here $n$ is either 0 or 1.
Hence, in the limit $\epsilon \to 0$ one has $\varphi\to 0$ and the second term is also vanishing. That is true even for the case when $A=B=0$.

Now consider the case when both points are on the mirror $x=y=-\beta t$; for such $x$ and $y$ we have $A=B=0$. We will be accurately taking the limit $A,B\to0$.
Using the limit $K_{1}(z)\to \frac{\Gamma(1)}{2}\frac{2}{z}=\frac{1}{z}$, as $z\to 0$, we obtain that
$$2m\epsilon\Big[\frac{K_{1}(m\sqrt{A^2+(iB+\epsilon)^2})}{\sqrt{A^2+(iB+\epsilon)^2}}+\frac{K_{1}(m\sqrt{A^2+(\epsilon-iB)^2})}{\sqrt{A^2+(\epsilon-iB)^2}}\Big]+2iBm\Big[\frac{K_{1}(m\sqrt{A^2+(iB+\epsilon)^2})}{\sqrt{A^2+(iB+\epsilon)^2}}-\frac{K_{1}(m\sqrt{A^2+(\epsilon-iB)^2})}{\sqrt{A^2+(\epsilon-iB)^2}}\Big]\to$$$$
2m\epsilon\Big[\frac{1}{m(A^2+(iB+\epsilon)^2)}+\frac{1}{m(A^2+(\epsilon-iB)^2)}\Big]+2iBm\Big[\frac{1}{m(A^2+(iB+\epsilon)^2)}-\frac{1}{m(A^2+(\epsilon-iB)^2)}\Big]=$$$$=
4\epsilon\frac{A^2+B^2+\epsilon^2}{(A^2-B^2+2iB\epsilon)(A^2-B^2-2iB\epsilon)}=\frac{2(A^2+B^2)}{B}\frac{\epsilon}{(A^2-B^2)^2+\epsilon^2}=\frac{2(A^2+B^2)}{B}\pi\delta(A^2-B^2)=$$$$2\pi\Big[\delta(A+B)+\delta(A-B)\Big]$$
Thus, in such a case in the commutation relations we also obtain the boundary contribution to the RHS, which are not zero only on the mirror.

Finally, the commutation relations are:
\begin{equation}
\label{CR}
[\phi(t,x),\pi(t,y)]=i\Big\{\delta\big(x-y\big)-\frac{1}{2}\delta\big[2\gamma^2\beta(1-\beta)t+
(1-\beta)^2\gamma^2x+y\big]-\frac{1}{2}\delta\big[2\gamma^2\beta(1+\beta)t+
(1+\beta)^2\gamma^2x+y\big]\Big\}.
\end{equation}

\subsection{The free Hamiltonian}

The proper Hamiltonian for massive field in the presence of moving mirror is:
$$H_{proper}=\frac{1}{2}\int\limits_{-\beta t}^{+\infty}\Big[\pi^2+(\partial_{x}\phi)^2+m^2\phi^2\Big]dx-\frac{\beta}{2}\Big\{\int\limits_{-\beta t}^{+\infty}\pi(x)\partial_{x}\phi(x) dx +\int\limits_{-\beta t}^{+\infty}\partial_{x}\phi(x)\pi(x) dx\Big\}.$$
As was explained in the case of massless fields above the proper Hamiltonian should has the following form: $H_{proper}=H-\beta P$, where $P$ is the momentum operator and $H$ is the standard Hamiltonian. Using the equations of motion we can rewrite it as follows:
\begin{equation}
H=\frac{1}{2}\int\limits_{-\beta t}^{+\infty}\Big[(\partial_{t}\phi)^2-\phi\partial_{t}^2\phi\Big]dx+\frac{1}{2}\phi\partial_{x}\phi\Big|^{\infty}_{-\beta t}-\beta P=\frac{1}{2}\int\limits_{-\beta t}^{+\infty}\Big[(\partial_{t}\phi)^2-\phi\partial_{t}^2\phi\Big]dx-\beta P,
\end{equation}
where

$$
P=\frac{1}{2}\int\limits_{-\beta t}^{+\infty}\Big[\partial_{t}\phi\partial_{x}\phi+\partial_{x}\phi\partial_{t}\phi\Big]dx.
$$
The operators that appear in these expressions are:
$$
\phi(t,x)=i\int\limits_{\gamma\beta m}^{+\infty} \frac{dk}{2\pi}\frac{1}{\sqrt{2\omega}}\Big[a_{k}\Big(e^{-i\omega t-ikx}-e^{-i\omega_{r} t+ik_{r}x}\Big)-a_{k}^\dagger\Big(e^{i\omega t+ikx}-e^{i\omega_{r} t-ik_{r}x}\Big)\Big],
$$
$$
\partial_{x}\phi(t,x)=\int\limits_{\gamma\beta m}^{+\infty} \frac{dk}{2\pi}\frac{1}{\sqrt{2\omega}}\Big[a_{k}\Big(k e^{-i\omega t-ikx}+k_{r}e^{-i\omega_{r}t+ik_{r} x}\Big)+a_{k}^\dagger\Big(ke^{i\omega t+ikx}+k_{r}e^{i\omega_{r} t-ik_{r}x}\Big)\Big],
$$
$$
\partial_{t}\phi(t,x)=\int\limits_{\gamma\beta m}^{+\infty} \frac{dk}{2\pi}\frac{1}{\sqrt{2\omega}}\Big[a_{k}\Big(\omega e^{-i\omega t-ikx}-\omega_{r}e^{-i\omega_{r}t+ik_{r} x}\Big)+a_{k}^\dagger\Big(\omega e^{i\omega t+ikx}-\omega_{r}e^{i\omega_{r} t-ik_{r}x}\Big)\Big],
$$
$$
\partial_{t}^{2}\phi(t,x)=-i \int\limits_{\gamma\beta m}^{+\infty} \frac{dk}{2\pi}\frac{1}{\sqrt{2\omega}}\Big[a_{k}\Big(\omega^2 e^{-i\omega t-ikx}-\omega^2_{r}e^{-i\omega_{r}t+ik_{r} x}\Big)-a_{k}^\dagger\Big(\omega^2 e^{i\omega t+ikx}-\omega^2_{r}e^{i\omega_{r} t-ik_{r}x}\Big)\Big].
$$
Now we will calculate different contributions to the Hamiltonian separately.

\subsubsection{Non-diagonal $a_{k}a_{k'}$ and $a^{\dagger}_{k}a^{\dagger}_{k'}$ parts}

Consider the contributions to the $a_{k}a_{k'}$ terms:
$$\frac{1}{4}\int\limits_{-\beta t}^{+\infty}dx\int\limits_{\gamma\beta m}^{+\infty}\int\limits_{\gamma\beta m}^{+\infty}\frac{dk}{2\pi}\frac{dk'}{2\pi}a_{k}a_{k'}\frac{\omega\omega'-\omega'^2-\beta(\omega k'+k\omega')}{\sqrt{\omega\omega'}}e^{-i(\omega+\omega')t-i(k+k')x}+$$
$$+\frac{1}{4}\int\limits_{-\beta t}^{+\infty}dx\int\limits_{\gamma\beta m}^{+\infty}\int\limits_{\gamma\beta m}^{+\infty}\frac{dk}{2\pi}\frac{dk'}{2\pi}a_{k}a_{k'}\frac{-\omega\omega'_{r}+\omega'^2_{r}-\beta(\omega k'_{r}-k\omega'_{r})}{\sqrt{\omega\omega'}}e^{-i(\omega+\omega_{r}')t+i(k_{r}'-k)x}+$$
$$+\frac{1}{4}\int\limits_{-\beta t}^{+\infty}dx\int\limits_{\gamma\beta m}^{+\infty}\int\limits_{\gamma\beta m}^{+\infty}\frac{dk}{2\pi}\frac{dk'}{2\pi}a_{k}a_{k'}\frac{-\omega_{r}\omega'+\omega'^2-\beta(k_{r}\omega'-\omega_{r}k')}{\sqrt{\omega\omega'}}e^{-i(\omega'+\omega_{r})t+i(k_{r}-k')x}+$$
$$+\frac{1}{4}\int\limits_{-\beta t}^{+\infty}dx\int\limits_{\gamma\beta m}^{+\infty}\int\limits_{\gamma\beta m}^{+\infty}\frac{dk}{2\pi}\frac{dk'}{2\pi}a_{k}a_{k'}\frac{\omega_{r}\omega'_{r}-\omega'^2_{r}+\beta(\omega_{r}k'_{r}+k_{r}\omega'_{r})}{\sqrt{\omega\omega'}}e^{-i(\omega_{r}+\omega_{r}')t+i(k_{r}+k'_{r})x}$$
Integrating over $x$ and using $\omega -\beta k=\omega_{r}+\beta k_{r}$, one can see that there is the same exponent $e^{-i(\omega+\omega')t+i\beta(k+k')t}$ in every member of the last sum. After the integration over $x$ we obtain the following expression multiplying the exponent:

\begin{equation}
\label{sts}
\begin{gathered}
\frac{\omega\omega'-\omega'^2-\beta(\omega k'+k\omega')}{k+k'-i\epsilon}-\frac{-\omega\omega'_{r}+\omega'^2_{r}-\beta(\omega k'_{r}-k\omega'_{r})}{k'_{r}-k+i\epsilon}-\frac{-\omega_{r}\omega'+\omega'^2-\beta(k_{r}\omega'-\omega_{r}k')}{k_{r}-k'+i\epsilon}-\\
-\frac{\omega_{r}\omega'_{r}-\omega'^2_{r}+\beta(\omega_{r}k'_{r}+k_{r}\omega'_{r})}{k_{r}+k'_{r}+i\epsilon},
\end{gathered}
\end{equation}
where we have introduced the following regularization: $e^{i(\omega'-\omega)t+i(k'-k)x} \to e^{i(\omega'-\omega)t+i(k'-k+i\epsilon)x}$.

Below we will use the following relations:
$$
\omega-\omega_{r}=2\beta\gamma^2(k-\beta\omega),
$$
$$
\omega+\omega_{r}=2\gamma^2(\omega-\beta k),
$$
$$
k\omega_{r}-\omega k_{r}=2\gamma^2\beta m^2,
$$
$$
k\omega_{r}+\omega k_{r}=2\gamma^2(\omega-\beta k)(k-\beta\omega),
$$
$$
k-k_{r}=2\beta\gamma^2(\omega-\beta k),
$$
$$
k+k_{r}=2\gamma^2(k-\beta\omega).
$$
Then, it is straightforward to show that in $\eqref{sts}$ we have:
$$Sum_{1}=\frac{\omega\omega'-\omega'^2}{k+k'-i\epsilon}-\frac{-\omega\omega'_{r}+\omega'^2_{r}}{k'_{r}-k+i\epsilon}=\frac{2\gamma^2\omega(\omega'-\beta k')(k'-\beta\omega')-2\beta\gamma^2\omega k(k'-\beta\omega')+4\beta\gamma^4k(k'-\beta\omega')(\omega'-\beta k')}{(k+k'-i\epsilon)(k'_{r}-k+i\epsilon)}-$$$$-\frac{\omega'^{2} k'_{r}+\omega'^{2}_{r} k'}{(k+k'-i\epsilon)(k'_{r}-k+i\epsilon)}+\frac{i\epsilon[2\beta\gamma^2(k'-\beta\omega')-4\beta\gamma^4(k'-\beta\omega')(\omega'-\beta k')]}{(k+k'-i\epsilon)(k'_{r}-k+i\epsilon)}.$$
Using $\delta(x)=\lim\limits_{\epsilon\to 0}\frac{\epsilon}{\pi(x^2+\epsilon^2)}$ and $x\delta(x)=0$, we conclude that
$$\lim_{\epsilon \to 0}\frac{i\epsilon\big[2\beta\gamma^2(k'-\beta\omega')-4\beta\gamma^4(k'-\beta\omega')(\omega'-\beta k')\big]}{(k+k'-i\epsilon)(k'_{r}-k+i\epsilon)}= $$$$=\frac{i\big[2\beta\gamma^2(k'-\beta\omega')-4\beta\gamma^4(k'-\beta\omega')(\omega'-\beta k')\big](k+k')(k'_{r}-k)\delta(k+k')}{(k'_{r}-k)^2}=0.$$
Similarly one can show that all terms in the sum under discussion that contain $\epsilon$ in the enumerator do vanish. Hence, below we do not show them.

Then one can show that in $\eqref{sts}$ we have:
$$Sum_{2}=-\frac{-\omega_{r}\omega'+\omega'^2}{k_{r}-k'+i\epsilon}-\frac{\omega_{r}\omega'_{r}-\omega'^2_{r}}{k_{r}+k'_{r}+i\epsilon}=\frac{2\gamma^2\omega_{r}(\omega'-\beta k')(k'-\beta\omega')+2\beta\gamma^2\omega_{r} k_{r}(k'-\beta\omega')-4\beta\gamma^4k_{r}(k'-\beta\omega)(\omega'-\beta k')}{(k_{r}-k'+i\epsilon)(k_{r}+k'_{r}+i\epsilon)}-$$$$ -\frac{\omega'^{2} k'_{r}+\omega'^{2}_{r} k'}{(k_{r}-k'+i\epsilon)(k_{r}+k'_{r}+i\epsilon)},$$
and
$$Sum_{3}=\beta\Big[\frac{\omega k_{r}'-k\omega_{r}'}{k_{r}'-k+i\epsilon}-\frac{\omega k'+k\omega'}{k+k'-i\epsilon}\Big]=\beta\frac{2\gamma^2\omega k(k'-\beta\omega')+2\beta\gamma^2k^2(k'-\beta\omega')-2\gamma^2k(k'-\beta\omega')(\omega'-\beta k')}{(k+k'-i\epsilon)(k+k'-i\epsilon)},$$
and
$$Sum_{4}=\beta\Big[\frac{\omega' k_{r}-k'\omega_{r}}{k_{r}-k'+i\epsilon}-\frac{\omega_{r} k_{r}'+k_{r}\omega_{r}'}{k_{r}+k_{r}'+i\epsilon}\Big]=\beta\frac{-2\gamma^2\omega_{r} k_{r}(k'-\beta\omega')+2\beta\gamma^2k_{r}^{2}(k'-\beta\omega')+2\gamma^2k_{r}(k'-\beta\omega)(\omega'-\beta k')}{(k_{r}-k'+i\epsilon)(k_{r}+k_{r}'+i\epsilon)}.$$
Then,
$$Sum_{1}+Sum_{3}=\frac{2\beta^2\gamma^2k^2(k'-\beta\omega')+(\omega-\beta k+2\beta\gamma^2k)2\gamma^2(k'-\beta\omega')(\omega'-\beta k')}{(k+k'-i\epsilon)(k'_{r}-k+i\epsilon)}-\frac{\omega'^{2} k'_{r}+\omega'^{2}_{r} k'}{(k+k'-i\epsilon)(k'_{r}-k+i\epsilon)},$$
and
$$Sum_{2}+Sum_{4}=\frac{2\beta^2\gamma^2k_{r}^2(k'-\beta\omega')+(\omega_{r}+\beta k_{r}-2\beta\gamma^2k_{r})2\gamma^2(k'-\beta\omega')(\omega'-\beta k')}{(k_{r}-k'+i\epsilon)(k_{r}+k'_{r}+i\epsilon)}-\frac{\omega'^{2} k'_{r}+\omega'^{2}_{r} k'}{(k_{r}-k'+i\epsilon)(k_{r}+k'_{r}+i\epsilon)}.$$
Using that:
$$\omega'^{2} k'_{r}+\omega'^{2}_{r} k'=2\gamma^2(k'-\beta\omega')(\omega'^2-\beta k'\omega'-\beta k'\omega'_{r})= \\ \\ =2\gamma^2(k'-\beta\omega')\big[\omega'^2-2\beta\gamma^2k'(\omega'-\beta k')\big],$$
we obtain that:
$$Sum_{2}+Sum_{4}=2\gamma^2(k'-\beta\omega')\frac{\beta^2k_{r}^2+(\omega_{r}+\beta k_{r}-2\beta\gamma^2k_{r})(\omega'-\beta k')-[\omega'^2-2\beta\gamma^2k'(\omega'-\beta k')]}{(k_{r}-k'+i\epsilon)(k_{r}+k'_{r}+i\epsilon)},$$
and
$$Sum_{1}+Sum_{3}=2\gamma^2(k'-\beta\omega')\frac{\beta^2k^2+(\omega-\beta k+2\beta\gamma^2k)(\omega'-\beta k')-[\omega'^2-2\beta\gamma^2k'(\omega'-\beta k')]}{(k+k'-i\epsilon)(k'_{r}-k+i\epsilon)}.$$
Hence,
$$Sum_{1}+Sum_{2}+Sum_{3}+Sum_{4}=2\gamma^2(k'-\beta\omega')\bigg[\frac{\beta^2k_{r}^2}{(k_{r}-k'+i\epsilon)(k_{r}+k'_{r}+i\epsilon)}+\frac{\beta^2k^2}{(k+k'-i\epsilon)(k'_{r}-k+i\epsilon)}+$$$$+\Big((\omega-\beta k)(\omega'-\beta k')-\omega'^2+2\beta\gamma^2k'(\omega'-\beta k')\Big)\Big(\frac{1}{(k_{r}-k'+i\epsilon)(k_{r}+k'_{r}+i\epsilon)}+\frac{1}{(k+k'-i\epsilon)(k'_{r}-k+i\epsilon)}\Big)+$$$$+
2\beta\gamma^2(\omega'-\beta k')\Big(\frac{k}{(k_{r}-k'+i\epsilon)(k_{r}+k'_{r}+i\epsilon)}-\frac{k_{r}}{(k+k'-i\epsilon)(k'_{r}-k+i\epsilon)}\Big)\bigg].$$
After the transformation of the last expression into one fraction, its enumerator has the following form:
$$Enumerator=8\beta\gamma^6(k'-\beta\omega')\bigg[-\beta^2(k-\beta\omega)\Big(kk_{r}(\omega'-\beta k')+k'k_{r}'(\omega-\beta k)\Big)- $$$$ -(k-\beta\omega)(\omega-\beta k+\omega'-\beta k')\Big((\omega-\beta k)(\omega'-\beta k')-\omega'^2+2\beta\gamma^2k'(\omega'-\beta k')\Big)+
(\omega'-\beta k')\Big((k-\beta\omega)(kk_{r}-k'k'_{r})\Big)\bigg].$$
We now show that this enumerator is equal to zero.
In this expression, we have:
$$kk_{r}-k'k'_{r}=\gamma^2(\omega-\beta k)^2-\gamma^2(\omega'-\beta k')^2=\gamma^2\Big(\omega-\beta k+\omega'-\beta k'\Big)\Big(\omega-\beta k-\omega'+\beta k'\Big),$$
$$kk_{r}(\omega'-\beta k')+k'k_{r}'(\omega-\beta k)=\gamma^2\Big(\omega-\beta k+\omega'-\beta k'\Big)\Big[(\omega-\beta k)(\omega'-\beta k')-m^2\Big],$$
where we have used that $kk_{r}=\gamma^2\Big((\omega-\beta k)^2-m^2\Big).$

Then we simplify the enumerator:
$$Enumerator=8\beta\gamma^6(k'-\beta\omega')(k-\beta\omega)(\omega-\beta k+\omega'-\beta k')\times $$$$ \times\Big[-\beta^2\gamma^2\Big((\omega-\beta k)(\omega'-\beta k')-m^2\Big)-(\omega-\beta k)(\omega'-\beta k')+\omega'^2-2\beta\gamma^2k'(\omega'-\beta k')+
\gamma^2(\omega'-\beta k')(\omega-\beta k-\omega'+\beta k')\Big]=$$
$$=8\beta\gamma^6(k'-\beta\omega')(k-\beta\omega)(\omega-\beta k+\omega'-\beta k') \Big(\beta^2\gamma^2m^2+\omega'^2-2\beta\gamma^2k'(\omega'-\beta k')-\gamma^2(\omega'-\beta k')^2\Big)=$$
$$=8\beta\gamma^8(k'-\beta\omega')(k-\beta\omega)(\omega-\beta k+\omega'-\beta k')(\omega'-\beta k')\Big(\omega'+\beta k'-2\beta k'-(\omega'-\beta k')\Big)=0,$$
where we have used that $m^2=\omega'^2-k'^2.$

So, the expression $\eqref{sts}$ is exactly zero. Thus, we have shown that $a_{k}a_{k'}$ and $a^\dagger_{k}a^\dagger_{k'}$ terms in Hamiltonian are vanishing.

\subsubsection{Diagonal $a_{k}a^\dagger_{k'}$ and $a^{\dagger}_{k}a_{k'}$ parts}

The $a_{k}a^\dagger_{k'}$ terms, in the Hamiltonian are as follows:
\begin{equation}
\label{sta}
\begin{gathered}
\frac{1}{4}\int\limits_{-\beta t}^{+\infty}dx\int\limits_{\gamma\beta m}^{+\infty}\int\limits_{\gamma\beta m}^{+\infty}\frac{dk}{2\pi}\frac{dk'}{2\pi}a_{k}a^\dagger_{k'}\frac{\omega\omega'+\omega'^2-\beta(\omega k'+k\omega')}{\sqrt{\omega\omega'}}e^{i(\omega'-\omega)t+i(k'-k)x}+\\+\frac{1}{4}\int\limits_{-\beta t}^{+\infty}dx\int\limits_{\gamma\beta m}^{+\infty}\int\limits_{\gamma\beta m}^{+\infty}\frac{dk}{2\pi}\frac{dk'}{2\pi}a_{k}a^\dagger_{k'}\frac{-\omega\omega'_{r}-\omega'^2_{r}-\beta(\omega k'_{r}-k\omega'_{r})}{\sqrt{\omega\omega'}}e^{i(\omega'_{r}-\omega_{r})t-i(k_{r}'+k)x}+\\+\frac{1}{4}\int\limits_{-\beta t}^{+\infty}dx\int\limits_{\gamma\beta m}^{+\infty}\int\limits_{\gamma\beta m}^{+\infty}\frac{dk}{2\pi}\frac{dk'}{2\pi}a_{k}a^\dagger_{k'}\frac{-\omega_{r}\omega'-\omega'^2-\beta(k_{r}\omega'-\omega_{r}k')}{\sqrt{\omega\omega'}}e^{i(\omega'-\omega_{r})t+i(k_{r}+k')x}+\\+\frac{1}{4}\int\limits_{-\beta t}^{+\infty}dx\int\limits_{\gamma\beta m}^{+\infty}\int\limits_{\gamma\beta m}^{+\infty}\frac{dk}{2\pi}\frac{dk'}{2\pi}a_{k}a^\dagger_{k'}\frac{\omega_{r}\omega'_{r}+\omega'^2_{r}+\beta(\omega_{r}k'_{r}+k_{r}\omega'_{r})}{\sqrt{\omega\omega'}}e^{i(\omega'_{r}-\omega_{r})t-i(k'_{r}-k_{r})x}.
\end{gathered}
\end{equation}
We have to introduce a regularization into this expression. In the massless case we have introduced the regularization as follows: $e^{i\Omega(t-x)(k'-k)} \to e^{i\Omega(t-x)(k'-k+i\epsilon)}$, where $\Omega=\frac{1-\beta}{1+\beta}$. Hence in the massive case we also introduce a similar regularization: $e^{i(\omega'_{r}-\omega_{r})t-i(k'_{r}-k_{r})x} \to e^{i(\omega'_{r}-\omega_{r})t-i(k'_{r}-k_{r}-i\Omega\epsilon)x}$. In such a case the Hamiltonian in the massive case will reduce to the massless one in the limit $m\to0$.

After the integration over $x$ and use of $\omega -\beta k=\omega_{r}+\beta k_{r}$ we obtain the same exponent $e^{i(\omega'-\omega)t-i\beta(k'-k)t}$ in each member of the sum in $\eqref{sta}$. Also after the integration over $x$ one has the following sum multiplying the exponent:
\begin{equation}
\label{sth}
\begin{gathered}
Sum=-i\bigg[-\frac{\omega\omega'+\omega'^2-\beta(\omega k'+k\omega')}{k'-k+i\epsilon}+\frac{-\omega\omega'_{r}-\omega'^2_{r}-\beta(\omega k'_{r}-k\omega'_{r})}{k'_{r}+k-i\epsilon}-\\-\frac{-\omega_{r}\omega'-\omega'^2-\beta(k_{r}\omega'-\omega_{r}k')}{k_{r}+k'+i\epsilon}+\frac{\omega_{r}\omega'_{r}+\omega'^2_{r}+\beta(\omega_{r}k'_{r}+k_{r}\omega'_{r})}{k'_{r}-k_r-i\Omega\epsilon}e^{(\Omega-1)\epsilon\beta t}\bigg].
\end{gathered}
\end{equation}
In the exponent in the last term here one can safely put $\epsilon = 0$.

Below for convenience we split the computation of the expression above into several steps. Then in the expression under consideration there are terms as follows:

$$iSum_{1}=-\frac{\omega\omega'+\omega'^2}{k'-k+i\epsilon}+\frac{-\omega\omega'_{r}-\omega'^2_{r}}{k'_{r}+k-i\epsilon}=-\frac{2\gamma^2\omega(\omega'-\beta k')(k'-\beta\omega')+2\beta\gamma^2\omega k(k'-\beta\omega')+4\beta\gamma^4k(k'-\beta\omega')(\omega'-\beta k')}{(k'-k+i\epsilon)(k'_{r}+k-i\epsilon)}-$$$$-\frac{2\gamma^2(k'-\beta\omega')\Big(\omega'^2-2\beta\gamma^2k'(\omega'-\beta k')\Big)}{(k'-k+i\epsilon)(k'_{r}+k-i\epsilon)}+\frac{i\epsilon\Big[\omega(\omega'-\omega_{r}')+\omega'^2-\omega_{r}'^2\Big]}{(k'-k+i\epsilon)(k'_{r}+k-i\epsilon)}.$$
Further one should pay attention to the members of the sum containing $i\epsilon$ in the enumerator, because exactly these terms will give non-zero diagonal contributions to the Hamiltonian. Also in $\eqref{sth}$ there are contributions as follows:
$$iSum_{2}=\frac{\omega_{r}\omega'+\omega'^2}{k_{r}+k'+i\epsilon}+\frac{\omega_{r}\omega'_{r}+\omega'^2_{r}}{k'_{r}-k_{r}-i\Omega\epsilon}=\frac{2\gamma^2\omega_{r}(\omega'-\beta k')(k'-\beta\omega')-2\beta\gamma^2\omega_{r} k_{r}(k'-\beta\omega')-4\beta\gamma^4k_{r}(k'-\beta\omega')(\omega'-\beta k')}{(k_{r}+k'+i\epsilon)(k'_{r}-k_{r}-i\Omega\epsilon)}+$$$$+\frac{2\gamma^2(k'-\beta\omega')\Big(\omega'^2-2\beta\gamma^2k'(\omega'-\beta k')\Big)}{(k_{r}+k'+i\epsilon)(k'_{r}-k_{r}-i\Omega\epsilon)}-\frac{i\epsilon\Big[\omega_{r}(\omega_{r}'-\Omega\omega')+\omega_{r}'^2-\Omega\omega'^2\Big]}{(k_{r}+k'+i\epsilon)(k'_{r}-k_{r}-i\Omega\epsilon)};$$
$$iSum_{3}=\beta\bigg[-\frac{\omega k_{r}'-k\omega_{r}'}{k_{r}'+k-i\epsilon}+\frac{\omega k'+k\omega'}{k'-k+i\epsilon}\bigg]=\beta\frac{2\gamma^2\omega k(k'-\beta\omega')+2\beta\gamma^2k^2(k'-\beta\omega')+2\gamma^2k(k'-\beta\omega')(\omega'-\beta k')}{(k_{r}'+k-i\epsilon)(k'-k+i\epsilon)}- $$$$-\frac{i\epsilon\beta\Big[k(\omega'-\omega_{r}')+\omega(k'+k_{r}')\Big]}{(k_{r}'+k-i\epsilon)(k'-k+i\epsilon)};$$
$$iSum_{4}=\beta\bigg[\frac{\omega' k_{r}-k'\omega_{r}}{k_{r}+k'+i\epsilon}+\frac{\omega_{r} k_{r}'+k_{r}\omega_{r}'}{k'_{r}-k_{r}-i\epsilon}\bigg]=\beta\frac{2\gamma^2\omega_{r} k_{r}(k'-\beta\omega')-2\beta\gamma^2k_{r}^{2}(k'-\beta\omega')+2\gamma^2k_{r}(k'-\beta\omega')(\omega'-\beta k')}{(k_{r}+k'+i\epsilon)(k'_{r}-k_{r}-i\epsilon)}+ $$$$+\frac{i\epsilon2\beta\Big[k_{r}(\omega_{r}'-\Omega\omega')+\omega_{r}(k_{r}'+\Omega k)\Big]}{(k_{r}+k'+i\epsilon)(k'_{r}-k_{r}-i\Omega\epsilon)}.$$
It can be shown that in each sum $Sum_{1,2,3,4}$ after the multiplication of the enumerator by the expression which is complex conjugate to the denominator and combining all the terms that are proportional to $\epsilon$, one will obtain expressions which have the form $x \delta(x) = 0$ in the limit $\epsilon \to 0$. Hence, one can drop such terms.

Then, we simplify the rest:
$$i(Sum_{1}+Sum_{3})=2\gamma^2(k'-\beta\omega')\frac{(\omega'-\beta k')(\beta k'-\omega-2\beta\gamma^2k+2\beta\gamma^2k')+\beta^2k^2-\omega'^2}{(k'-k+i\epsilon)(k'_{r}+k-i\epsilon)},$$
$$i(Sum_{2}+Sum_{4})=2\gamma^2(k'-\beta\omega')\frac{(\omega'-\beta k')(\beta k_{r}+\omega_{r}-2\beta\gamma^2k_{r}-2\beta\gamma^2k_{r}')-\beta^2k_{r}^2+\omega'^2}{(k'+k_{r}+i\epsilon)(k'_{r}-k_{r}-i\Omega\epsilon)},$$
and then the full sum is as follows:
\begin{equation}
\label{stj}
\begin{gathered}
iSum=i(Sum_{1}+Sum_{2}+Sum_{3}+Sum_{4})=8\beta\gamma^6(k'-\beta\omega')(k-\beta\omega)(\omega-\beta k-\omega'+\beta k')\times \\ \times\frac{-\beta^2\gamma^2\Big((\omega-\beta k)(\omega'-\beta k')+m^2\Big)+\gamma(\omega-\beta k+\omega'-\beta k')(\omega'-\beta k')-(\omega'-\beta k')(\omega-\beta k-2\beta\gamma^2k')-\omega'^2}{(k'-k+i\epsilon)(k'_{r}+k-i\epsilon)(k'+k_{r}+i\epsilon)(k'_{r}-k_{r}-i\Omega\epsilon)}+\\+
8\gamma^4(k'-\beta\omega')(k-\beta\omega)i\epsilon\times \\ \times \frac{(\omega-\beta k)(\omega'-\beta k')-2\beta\gamma^2k'(\omega'-\beta k')+\omega'^2-\beta^2\Big[2\beta^2\gamma^4(\omega-\beta k)(\omega'-\beta k')+(\omega-\beta k)^2-m^2\gamma^2\Big]+2\beta^2\gamma^4(\omega'-\beta k')^2}{(k'-k+i\epsilon)(k'_{r}+k-i\epsilon)(k'+k_{r}+i\epsilon)(k'_{r}-k_{r}-i\Omega\epsilon)}+\\
+2\gamma^2(k'-\beta\omega')(1-\Omega)i\epsilon \times \\ \times\frac{-\Big[(\omega-\beta k)(\omega'-\beta k')-2\beta\gamma^2k'(\omega'-\beta k')+\omega'^2\Big](k'+k_{r})+\beta^2(k'k^2+k_{r}k^2)-2\beta\gamma^2(\omega'-\beta k')(kk_{r}+kk')}{(k'-k+i\epsilon)(k'_{r}+k-i\epsilon)(k'+k_{r}+i\epsilon)(k'_{r}-k_{r}-i\Omega\epsilon)}.
\end{gathered}
\end{equation}
The enumerator of the first contribution to the sum in the last expression is almost the same as in $a_{k}a_{k'}$ term. Then,
$$-\beta^2\gamma^2\Big((\omega-\beta k)(\omega'-\beta k')+m^2\Big)+\gamma(\omega-\beta k+\omega'-\beta k')(\omega'-\beta k')-(\omega'-\beta k')(\omega-\beta k-2\beta\gamma^2k')-\omega'^2= $$$$=-\beta^2\gamma^2m^2+\gamma(\omega'-\beta k')^2+2\beta\gamma^2k'(\omega'-\beta k')-\omega'^2=\gamma^2(\omega'-\beta k')(\omega'+\beta k'-\omega'+\beta k'-2\beta k')=0$$
At the same time, the second contribution in $\eqref{stj}$ is
$$8\gamma^4(k'-\beta\omega')(k-\beta\omega)i\epsilon\times $$$$\times \frac{(\omega-\beta k)(\omega'-\beta k')-2\beta\gamma^2k'(\omega'-\beta k')+\omega'^2-\beta^2\Big[2\beta^2\gamma^4(\omega-\beta k)(\omega'-\beta k')+(\omega-\beta k)^2-m^2\gamma^2\Big]+2\beta^2\gamma^4(\omega'-\beta k')^2}{(k'-k+i\epsilon)(k'_{r}+k-i\epsilon)(k'+k_{r}+i\epsilon)(k'_{r}-k_{r}-i\Omega\epsilon)}= $$$$
8\gamma^4(k'-\beta\omega')(k-\beta\omega)i\pi\delta(k'-k)\times $$$$ \times\frac{\Big[(\omega-\beta k)^2(1-2\beta^4\gamma^4-\beta^2\gamma^2+2\beta^2\gamma^4)-2\beta\gamma^2k(\omega-\beta k)+\omega^2+m^2\gamma^2\beta^2\Big](k'_{r}+k)(k'+k_{r})(k'_{r}-k_{r})(k'-k)}{(k'_{r}+k)^2(k'+k_{r})^2(k'_{r}-k_{r})^2}= $$$$
8\gamma^6(k'-\beta\omega')(k-\beta\omega)\frac{\Big[(\omega-\beta k)^2-2\beta k(\omega-\beta k)+(\omega-\beta k)(\omega+\beta k)\Big](k'-k)}{(k'_{r}+k)(k'+k_{r})(k'_{r}-k_{r})}i\pi\delta(k'-k)= \\ \\4\gamma^2(\omega-\beta k)^2\frac{k'-k}{k'_{r}-k_{r}}i\pi\delta(k'-k),$$
where we have used that $\delta(x)=\lim\limits_{\epsilon\to0}\frac{\epsilon}{\pi(x^2+\epsilon^2)}$. The last expression can be transformed as
$$\frac{k'-k}{k'_{r}-k_{r}}\delta(k'-k)=\frac{k'-k}{(k'-k)\gamma^2(1+\beta^2)-2\beta\gamma^2(\omega'-\omega)}
\frac{\omega'+\omega}{\omega'+\omega}\delta(k'-k)= $$$$ \frac{(k'-k)(\omega'+\omega)}{\gamma^2(k'-k)\Big((1+\beta^2)(\omega'+\omega)-2\beta(k'+k)\Big)}\delta(k'-k)=\frac{\omega}{\omega_{r}}\delta(k'-k),$$
and the second member of the expression $\eqref{stj}$ is $4\gamma^2(\omega-\beta k)^2\frac{\omega}{\omega_{r}}i\pi\delta(k'-k)$.

Finally, the last member of the expression $\eqref{stj}$ is
$$2\gamma^2(k-\beta\omega)(\Omega-1)(k+k_{r})\frac{\Big[(\omega-\beta k)^2+\omega^2-\beta^2k^2\Big](k'_{r}+k)(k'+k_{r})(k'_{r}-k_{r})(k'-k)}{(k'_{r}+k)^2(k'+k_{r})^2(k'_{r}-k_{r})^2}i\pi\delta(k'-k)=$$$$
-\frac{2\beta}{1+\beta}(\omega-\beta k)2\omega\frac{\omega}{\omega_{r}}i\pi\delta(k'-k).$$
Combining all these terms together, we obtain that the Hamiltonian for the massive field in the presence of the mirror is
\begin{equation}\label{34}
H=\int\limits_{\gamma\beta m}^{+\infty}\frac{dk}{2\pi}\frac{\gamma^2(\omega-\beta k)\Big(\omega-\beta k-\beta(1-\beta)\omega\Big)}{2\omega_{r}}\Big(a_{k}a^\dagger_{k}+a^\dagger_{k}a_{k}\Big),
\end{equation}
where $\omega_{r}=(1+\beta^2)\gamma^2\omega-2\beta\gamma^2k$.

One can see that when $k=\omega, m=0$ the Hamiltonian reduces to
$$H=(1-\beta)\int\limits_{0}^{+\infty}\frac{dk}{2\pi}\frac{k}{2}\Big(a_{k}a^\dagger_{k}+a^\dagger_{k}a_{k}\Big),$$
i.e. into the Hamiltonian for massless field.

Also note that on the lower limit of integration over $k$ the integrand in (\ref{34}) is equal to $\frac{\gamma^2(\omega-\beta k)(\omega-\beta k-\beta(1-\beta)\omega)}{2\omega_{r}}\Big|_{k=\gamma\beta m}=\frac{m}{2}\sqrt{\frac{1-\beta}{1+\beta}}$.

\subsection{The stress-energy flux}

Using the mode decomposition of the field operator, the vacuum expectation value of the flux can be represented as
$$\braket{0|T_{tx}|0}=\int\limits_{\gamma\beta m}^{+\infty} \frac{dk}{2\pi}ke^{-\omega \epsilon}-\int\limits_{\gamma\beta m}^{+\infty} \frac{dk}{2\pi}\frac{k_{r}\omega_{r}}{\omega}e^{-\omega_{r} \epsilon}-$$$$-\gamma^2\beta m^2\int\limits_{\gamma\beta m}^{+\infty} \frac{dk}{2\pi}\frac{e^{i(\omega_{r}-\omega)t-i(k_{r}+k)x-\omega_{r}\epsilon}}{\omega}-\gamma^2\beta m^2\int\limits_{\gamma\beta m}^{+\infty} \frac{dk}{2\pi}\frac{e^{-i(\omega_{r}-\omega)t+i(k_{r}+k)x-\omega\epsilon}}{\omega}.$$
Then, using $\frac{dk}{\omega}=\frac{dk_{r}}{\omega_{r}}$, we obtain
$$\braket{0|T_{tx}|0}=-\gamma^2\beta m^2\Bigl(\int\limits_{\gamma\beta m}^{+\infty} \frac{dk}{2\pi}\frac{e^{i(\omega_{r}-\omega)t-i(k_{r}+k)x-\omega_{r}\epsilon}}{\omega}+\int\limits_{\gamma\beta m}^{+\infty} \frac{dk}{2\pi}\frac{e^{-i(\omega_{r}-\omega)t+i(k_{r}+k)x-\omega\epsilon}}{\omega}\Bigr).$$
Consider the first integral in this expression. Make the change $\omega_{r}=\omega'$ and $ k_{r}=k'$, then $k=\gamma^2[(1+\beta^2)k'+2\beta\omega']$ and $\omega=\gamma^2[(1+\beta^2)\omega'+2\beta k']$.
One can rewrite this integral as:
$$\int\limits_{\gamma\beta m}^{+\infty} \frac{dk}{2\pi}\frac{e^{i(\omega_{r}-\omega)t-i(k_{r}+k)x-\omega_{r}\epsilon}}{\omega}=\int\limits_{-\gamma\beta m}^{+\infty} \frac{dk'}{2\pi}\frac{e^{i(\omega'-\gamma^2[(1+\beta^2)\omega'+2\beta k'])t-i(k'+\gamma^2[(1+\beta^2)k'+2\beta\omega'])x-\omega'\epsilon}}{\omega'}=$$ $$=\big\{k'\to -k'\big\}=\int\limits_{-\infty}^{\gamma\beta m} \frac{dk'}{2\pi}\frac{e^{-i(\omega'_{r}-\omega')t+i(k'_{r}+k')x-\omega'\epsilon}}{\omega'}=\int\limits_{-\infty}^{\gamma\beta m} \frac{dk}{2\pi}\frac{e^{-i(\omega_{r}-\omega)t+i(k_{r}+k)x-\omega\epsilon}}{\omega}.$$
As a result, we obtain the following expression:
\begin{equation}
\braket{0|T_{tx}|0}=-\gamma^2\beta m^2\int\limits_{-\infty}^{+\infty} \frac{dk}{2\pi}\frac{e^{i2\gamma^2(k-\beta\omega)(\beta t+x)-\omega\epsilon}}{\omega}.
\end{equation}
We rewrite it as
$$\int\limits_{-\infty}^{+\infty} \frac{dk}{\pi}\frac{e^{iAk-iB\sqrt{k^2+1}}}{2\sqrt{k^2+1}}=\int\limits_{0}^{+\infty} \frac{dk}{\pi}\frac{\cos(Ak)}{\sqrt{k^2+1}}e^{-iB\sqrt{k^2+1}}=$$$$
\int\limits_{0}^{+\infty} \frac{dk}{\pi}\frac{\cos(Ak)}{\sqrt{k^2+1}}\cos\Big(B\sqrt{k^2+1}\Big)+\int\limits_{0}^{+\infty} \frac{dk}{\pi}\frac{\cos(Ak)}{\sqrt{k^2+1}}\sin\Big(B\sqrt{k^2+1}\Big),$$
where we have changed $k\to mk$ and introduced the following notations: $A=2m\gamma^2(x+\beta t)$ and $B=2m\gamma^2\beta(x+\beta t-i\epsilon)$.

Using the table integrals:
\begin{equation}
\int\limits_{0}^{+\infty}dk\frac{\cos(Ak)}{\sqrt{k^2+1}}\cos\Big(B\sqrt{k^2+1}\Big)=K_{0}\Big(\sqrt{A^2-B^2}\Big),\quad {\rm  where }\quad A>|B|>0,
\end{equation}
\begin{equation}
\int\limits_{0}^{+\infty}dk\frac{\cos(Ak)}{\sqrt{k^2+1}}\sin\Big(B\sqrt{k^2+1}\Big)=0,\quad {\rm  where } \quad A>|B|>0,
\end{equation}
where $K_{0}$ is the modified Bessel Function of the second kind, we obtain that:
$$\braket{0|T_{tx}|0}=-\frac{1}{\pi}\gamma^2\beta m^2K_{0}\Big(2m\gamma^2\sqrt{(\beta t+x)^2-\beta^2(\beta t+x-i\epsilon)^2}\Big).$$
In the limit $\epsilon\to0$ this expression reduces to:
\begin{equation}
\label{flux}
\braket{0|T_{tx}|0}=-\frac{1}{\pi}\gamma^2\beta m^2K_{0}\big[2m\gamma(x+\beta t)\big].
\end{equation}
From here, one can see that at $x=-\beta t$, i.e. on the mirror, the vacuum expectation value is infinite, but when $x\to \infty$ the expectation value is exponentially decaying. We discuss the meaning of the obtained expression in the next subsection.

\subsubsection{Vacuum expectation values and the Lorentz invariance}

In this subsection we show that the above expression for $\braket{0|T_{tx}|0}$ can be obtained by boosting the mirror at rest.
First, we recall the calculation of the vacuum expectation value of the stress-energy tensor in the empty space. The mode expansion of the field operator in such a case is
$$
\phi(t,x)=\int\limits_{-\infty}^{+\infty}\frac{dk}{2\pi}\frac{1}{\sqrt{2\omega}}\Big(a_{k}e^{-i\omega t-ikx}+a_{k}^\dagger e^{i\omega t+ikx}\Big).
$$
Then the relevant vacuum expectation values are as follows:
$$\braket{T_{tt}}_{0}=\int\limits_{-\infty}^{+\infty}\frac{\omega}{2}\frac{dk}{2\pi},\quad{\rm and}\quad
\braket{T_{xx}}_{0}=\int\limits_{-\infty}^{+\infty}\frac{k^2}{2\omega}\frac{dk}{2\pi},\quad{\rm and}\quad
\braket{T_{tx}}_{0}=0,$$
where the subscript $\textquotedblright$ 0 $\textquotedblright$ means that the expectation value is taken in the empty space, i.e. without a mirror.
To obtain the Lorentz invariant expression for the expectation value we apply the following regularization procedure. We cut the momentum integration by $\Lambda$ (instead of that one can use the point-splitting regularization) and also subtract from it the same expression, but with a greater mass $M\gg m$. Such a regularization is inspired by the Pauli-Villars one.

The point is that the Pauli--Villars regularization alone does not cutoff all the divergences in the expectation values (e.g. it does not regulate the quadratic ones). At the same time the $\Lambda$ cutoff of the momentum integration violates the Lorentz invariance. At the same time the combination of those two allows to cut all the divergences and to respect the Lorentz invariance. That is because those terms which violate it cancel between the physical and Pauli--Villars fields.

For example, for the case of $\braket{T_{xx}}_{0}$ we obtain:
$$\braket{T_{xx}}_{0}=\int\limits_{-\Lambda}^{\Lambda}\frac{k^2}{2\sqrt{k^2+m^2}}\frac{dk}{2\pi}-\int\limits_{-\Lambda}^{\Lambda}\frac{k^2}{2\sqrt{k^2+M^2}}\frac{dk}{2\pi}=\frac{1}{4\pi}\Big[\Lambda^2\sqrt{1+\frac{m^2}{\Lambda^2}}-m^2\log{\Big(\sqrt{1+\frac{m^2}{\Lambda^2}}+1}\Big)\Big]-(m \to M)$$
So, in the limits $\Lambda \to \infty$ and  $M\gg m$ we find that
$$\braket{T_{xx}}_{0}=\frac{1}{4\pi}M^2\log{\Lambda}.$$
Similarly one can find the expression for $\braket{T_{tt}}_{0}$. As a result:
\begin{equation}
\braket{T_{\mu\nu}}_{0}=-\frac{1}{4\pi}\eta_{\mu\nu}M^2\log{\Lambda},
\end{equation}
where $\eta_{\mu\nu}=diag(1,-1)$ is the metric tensor. This answer is obviously Lorentz-invariant. In the presence of mirror the Lorentz invariance is obviously broken by it mere presence.

Let us repeat now the same computation in the presence of mirror at rest. First,
$$\braket{T_{tx}}=\int\limits_{0}^{+\infty}\frac{dk}{2\pi}(\sin{kx}\cos{kx}-\cos{kx}\sin{kx})=0.$$
Second,
$$\braket{T_{xx}}=2\int\limits_{0}^{\Lambda}\frac{\omega^2\sin^2(kx)+k^2\cos^2(kx)-m^2sin^2(kx)}{\omega}\frac{dk}{2\pi}-(m\rightarrow M)=2\int\limits_{0}^{\Lambda}\frac{k^2}{\omega}\frac{dk}{2\pi}-(m\rightarrow M)=2\braket{T_{xx}}_{0}.$$
Note that in the presence of mirror the range of integration
over $k$ is $[0,\Lambda]$, rather than $[-\Lambda,\Lambda]$.
Finally, to rederive the result $\eqref{flux}$ of the previous subsection in addition to the above regularizations we also need to use the point-splitting one:
$$\braket{T_{tt}}=2\int\limits_{0}^{\Lambda}\frac{\omega^2\sin^2(kx)+k^2\cos^2(kx)+m^2sin^2(kx)}{\omega}e^{-\omega\epsilon}\frac{dk}{2\pi}-(m\rightarrow M)=$$$$=2\braket{T_{tt}}_{0}-2\int\limits_{0}^{+\infty}\frac{m^2\cos(2kx)}{\omega}e^{-\omega\epsilon}\frac{dk}{2\pi}+2\int\limits_{0}^{+\infty}\frac{M^2\cos(2kx)}{\sqrt{k^2+M^2}}e^{-\sqrt{k^2+M^2}\epsilon}\frac{dk}{2\pi}=$$$$=2\braket{T_{tt}}_{0}-\frac{1}{\pi}m^2K_{0}(2mx)+\frac{1}{\pi}M^2K_{0}(2Mx)=2\braket{T_{tt}}_{0}-\frac{1}{\pi}m^2K_{0}(2mx),$$
where in second line we take the limits $\Lambda \to\infty$ and $M\to\infty$, but $1/\epsilon \gg M$.

Hence the normal ordered vacuum expectation value of the stress-energy tensor in the presence of mirror at rest is as follows:
\begin{equation}
\label{T00}
\braket{:T_{\mu\nu}:}=-\frac{1}{\pi}m^2K_{0}(2mx)\begin{pmatrix} 1&0\\
0&0\end{pmatrix}.
\end{equation}
The normal ordering is done by the subtraction from $\braket{T_{\mu\nu}}$ of the divergent expression that is proportional to $\eta_{\mu\nu}$, i.e. which is similar to the empty space one. The obtained expectation value violates the Lorentz invariance, which is natural in the presence of mirror.

Because of that, if one boosts the mirror at rest, he obtains the following vacuum expectation value of the flux:
$$\braket{T_{t'x'}}=\beta\gamma^2\Big[\braket{T_{tt}}+\braket{T_{xx}}\Big]=-\frac{1}{\pi}m^2\beta\gamma^2K_{0}(2mx)=-\frac{1}{\pi}m^2\beta\gamma^2K_{0}\Big[2m\gamma(x'+\beta t')\Big],$$
which coincides with the expression found in the previous subsection.

This non-zero answer for the vacuum expectation value of the stress-energy tensor in the presence of a moving mirror with constant velocity does not correspond to any flux. In fact, one can see that  $\braket{T_{t'x'}}$ has infinite value when $x=-\beta t$, but also rapidly decays to zero as $x\to \infty$. So one can think of the observed effect as if the mirror captures and carries along with itself a portion of zero-point fluctuations.

\section{Non-ideal mirror}

\subsection{Definition and modes}

The mirror we have considered above reflects all modes equally well. But a real mirror is transparent for modes with high enough momenta.
As a model of non-ideal mirror in this work we consider the delta-potential barrier $V(x)=\alpha\delta(x)$, where $\alpha$ is a dimensionfull coefficient. The K-G equation in such a case is
\begin{equation}
\label{eq:1}
\bigg(\partial^{2}_{t}-\partial^{2}_{x}+m^2+\alpha\delta(x)\bigg)h(t,x)=0, \quad \alpha>0.
\end{equation}
To find the modes we use the following sewing conditions:
\begin{equation}
\label{eq:2}
h(t,+0)=h(t,-0),
\end{equation}
\begin{equation}
\label{eq:3}
\partial_{x}h(t,+0)-\partial_{x}h(t,-0)=\alpha h(t,0).
\end{equation}
One can define the mode functions piecewise in different regions of space-time:
\begin{equation}
\label{eq:4}
h_{k>0}(t,x)=
\begin{cases}
\dfrac{e^{-i\omega t}}{\sqrt{2\omega}}\bigg(e^{-ikx}-\dfrac{\alpha}{2ik+\alpha}e^{ikx}\bigg) & x<0\\
\dfrac{e^{-i\omega t}}{\sqrt{2\omega}}\dfrac{2ik}{2ik+\alpha}e^{-ikx} &  x>0;
\end{cases}
\end{equation}
\begin{equation}
\label{eq:5}
h_{k<0}(t,x)=
\begin{cases}
\dfrac{e^{-i\omega t}}{\sqrt{2\omega}}\bigg(e^{-ikx}+\dfrac{\alpha}{2ik-\alpha}e^{ikx}\bigg) & x>0\\
\dfrac{e^{-i\omega t}}{\sqrt{2\omega}}\dfrac{2ik}{2ik-\alpha}e^{-ikx} &  x<0,
\end{cases}
\end{equation}
or with the use of the Heaviside function:
$$h(t,x)=\bigg[\theta(-x)\theta(-k)\dfrac{2ik}{2ik-\alpha}e^{-ikx}+\theta(-x)\theta(k)\Big(e^{-ikx}-\dfrac{\alpha}{2ik+\alpha}e^{ikx}\Big)+$$$$+\theta(x)\theta(-k)\Big(e^{-ikx}+\dfrac{\alpha}{2ik-\alpha}e^{ikx}\Big)+\theta(x)\theta(-k)\dfrac{2ik}{2ik+\alpha}e^{-ikx}\bigg]\dfrac{e^{-i\omega t}}{\sqrt{2\omega}}. $$
In the limit $\alpha\to0$ such modes reduce to their standard form: $\phi(t,x)=\dfrac{e^{-ikx-i\omega t}}{\sqrt{2\omega}}$. Note that for $k\gg\alpha$ the modes are not sensitive to the presence of the potential and look as regular plane waves, i.e. the mirror is transparent for them.

The quantized field is as follows:
\begin{equation}
\phi(t,x)=\int\limits_{-\infty}^{+\infty}\frac{dk}{2\pi}\frac{a_{k}h(x)e^{i\omega t}+a_{k}^\dagger h^{*}(x)e^{-i\omega t}}{\sqrt{2\omega}}.
\end{equation}
We show now that the commutation relations between the field and its conjugate momentum have the canonical form:
$$
-i[\phi(t,x),\partial_{t}\phi(t,y)]=$$$$\theta(-x)\theta(-y)\int\limits_{-\infty}^{+\infty}dk\bigg[\theta(-k)\Big|\frac{2ik}{2ik+\alpha}\Big|^2e^{ik(x-y)}+\theta(k)\Big(e^{ik(x-y)}+\Big|\frac{\alpha}{2ik-\alpha}\Big|^2e^{-ik(x-y)}-\frac{\alpha}{2ik+\alpha}e^{ik(x+y)}+\frac{\alpha}{2ik-\alpha}e^{-ik(x+y)}\Big)    \bigg]$$$$
+\theta(x)\theta(y)\int\limits_{-\infty}^{+\infty}dk\bigg[\theta(k)\Big|\frac{2ik}{2ik-\alpha}\Big|^2e^{ik(x-y)}+\theta(-k)\Big(e^{ik(x-y)}+\Big|\frac{\alpha}{2ik+\alpha}\Big|^2e^{-ik(x-y)}-\frac{\alpha}{2ik+\alpha}e^{-ik(x+y)}+\frac{\alpha}{2ik-\alpha}e^{ik(x+y)}\Big)    \bigg]$$$$
+\theta(-x)\theta(y)\int\limits_{-\infty}^{+\infty}dk\bigg[\theta(-k)\Big(\frac{2ik}{2ik+\alpha}e^{ik(x-y)}+\frac{2ik\alpha}{|2ik+\alpha|^2}e^{ik(x+y)}\Big)+\theta(k)\Big(\frac{2ik}{2ik+\alpha}e^{ik(x-y)}+\frac{2ik\alpha}{|2ik-\alpha|^2}e^{-ik(x-y)}\Big)\bigg]$$$$
+\theta(x)\theta(-y)\int\limits_{-\infty}^{+\infty}dk\bigg[\theta(-k)\Big(\frac{2ik}{2ik-\alpha}e^{ik(x-y)}-\frac{2ik\alpha}{|2ik+\alpha|^2}e^{-ik(x+y)}\Big)+\theta(k)\Big(\frac{2ik}{2ik-\alpha}e^{ik(x-y)}-\frac{2ik\alpha}{|2ik-\alpha|^2}e^{ik(x-y)}\Big)\bigg]=$$$$
=\theta(-x)\theta(-y)\int\limits_{-\infty}^{+\infty}dk\bigg[e^{ik(x-y)}-\frac{\alpha}{2ik+\alpha}e^{ik(x+y)}\bigg]+\theta(x)\theta(y)\int\limits_{-\infty}^{+\infty}dk\bigg[e^{ik(x-y)}+\frac{\alpha}{2ik-\alpha}e^{ik(x+y)}\bigg]$$$$
+\theta(-x)\theta(y)\int\limits_{-\infty}^{+\infty}dk\bigg[\frac{2ik}{2ik+\alpha}e^{ik(x-y)}\bigg]+\theta(x)\theta(-y)\int\limits_{-\infty}^{+\infty}dk\bigg[\frac{2ik}{2ik-\alpha}e^{ik(x-y)}\bigg],
$$
where we have used the change $k\to-k$ in the integrals.

It happens that most of the integrals above are vanishing. In fact,
$$\theta(x)\theta(y)\int\limits_{-\infty}^{+\infty}dk\bigg[\frac{\alpha}{2ik-\alpha}e^{ik(x+y)}\bigg]=0,$$
because the integrand in the analytic function in the lower plane. Also:
$$\theta(-x)\theta(y)\int\limits_{-\infty}^{+\infty}dk\bigg[\frac{2ik}{2ik+\alpha}e^{ik(x-y)}\bigg]=\theta(-x)\theta(y)\delta(x-y)+\theta(-x)\theta(y)\int\limits_{-\infty}^{+\infty}dk\bigg[\frac{\alpha}{2ik+\alpha}e^{ik(x-y)}\bigg]=0$$
Hence, we find the following commutation relations:
\begin{equation}
\label{eq:6}
[\phi(t,x),\partial_{t}\phi(t,y)]=i\Big(\theta(-x)\theta(-y)+\theta(x)\theta(y)\Big)\delta(x-y)=i\delta(x-y)
\end{equation}
Thus, unlike the case of the ideal mirror, in the present situation we do not find unusual terms in the commutation relations. That is one of the reasons why it is appropriate to consider non-ideal mirror rather than the ideal one.

\subsection{The free Hamiltonian}

Let us find the explicit form of the Hamiltonian operator via the creation and annihilation operators.  By definition we have that:
\begin{equation}
    \label{eq:7}
    H=\dfrac{1}{2}\int\limits_{-\infty}^{+\infty}dx \Big[(\partial_{t}\phi)^2+(\partial_{x}\phi)^2 +\Big(m^2+\alpha \delta(x)\Big)\phi^2\Big]=
\end{equation}
$$=\dfrac{1}{2}\int\limits_{-\infty}^{+\infty}dx \Big[(\partial_{t}\phi)^2+(\partial_{x}\phi)^2 +m^2 \phi^{2}+\alpha\phi^{2}(t,0)\Big].$$
Consider every term here separately:
\begin{multline}
\label{eq:s8}
    \int\limits_{-\infty}^{+\infty}dx(\partial_{t}\phi)^2=\int\limits_{-\infty}^{0}dx(\partial_{t}\phi)^2+\int\limits_{0}^{+\infty}dx(\partial_{t}\phi)^2= \\
    =\int\limits_{-\infty}^{0}dx\Bigg[\Bigg(\int\limits_{-\infty}^{0}\dfrac{dk}{2\pi}\bigg(\dfrac{-i\omega_{k}}{\sqrt{2\omega_{k}}}e^{-i\omega_{k}t}\dfrac{2ik}{2ik-\alpha}e^{-ikx}a_{k}+\dfrac{i\omega_{k}}{\sqrt{2\omega_{k}}}e^{i\omega_{k}t}\dfrac{2ik}{2ik+\alpha}e^{ikx}a^{\dag}_{k}\bigg)+\\
    +\int\limits_{0}^{+\infty}\dfrac{dk}{2\pi}\bigg(\dfrac{-i\omega_{k}}{\sqrt{2\omega_{k}}}e^{-i\omega_{k}t}\big(e^{-ikx}-\dfrac{\alpha}{2ik+\alpha}e^{ikx}\big)a_{k}+\dfrac{i\omega_{k}}{\sqrt{2w_{k}}}e^{i\omega_{k}t}\big(e^{ikx}+\dfrac{\alpha}{2ik-\alpha}e^{-ikx}\big)a_{k}^{\dag}\bigg)\Bigg)\times \\
    \times\Bigg(\int\limits_{-\infty}^{0}\dfrac{dq}{2\pi}\bigg(\dfrac{-i\omega_{q}}{\sqrt{2\omega_{q}}}e^{-i\omega_{q}t}\dfrac{2iq}{2iq-\alpha}e^{-iqx}a_{q}+\dfrac{i\omega_{q}}{\sqrt{2\omega_{q}}}e^{i\omega_{q}t}\dfrac{2iq}{2iq+\alpha}e^{iqx}a^{\dag}_{q}\bigg)+ \\
    +\int\limits_{0}^{+\infty}\dfrac{dq}{2\pi}\bigg(\dfrac{-i\omega_{q}}{\sqrt{2\omega_{q}}}e^{-i\omega_{q}t}\big(e^{-iqx}-\dfrac{\alpha}{2iq+\alpha}e^{iqx}\big)a_{q}+\dfrac{i\omega_{q}}{\sqrt{2\omega_{q}}}e^{i\omega_{q}t}\big(e^{iqx}+\dfrac{\alpha}{2iq-\alpha}e^{-iqx}\big)a_{q}^{\dag}\bigg)\Bigg)\Bigg] + \\
    +\int\limits_{0}^{+\infty}dx\Bigg[\Bigg(\int\limits_{-\infty}^{0}\dfrac{dk}{2\pi}\bigg(\dfrac{-i\omega_{k}}{\sqrt{2\omega_{k}}}e^{-i\omega_{k}t}\big(e^{-ikx}+\dfrac{\alpha}{2ik-\alpha}e^{ikx}\big)a_{k}+\dfrac{i\omega_{k}}{\sqrt{2\omega_{k}}}e^{i\omega_{k}t}\big(e^{ikx}-\dfrac{\alpha}{2ik+\alpha}e^{-ikx}\big)a_{k}^{\dag}\bigg)+ \\
    +\int\limits_{0}^{+\infty}\dfrac{dk}{2\pi}\bigg(\dfrac{-i\omega_{k}}{\sqrt{2\omega_{k}}}e^{-i\omega_{k}t}\dfrac{2ik}{2ik+\alpha}e^{-ikx}a_{k}+\dfrac{i\omega_{k}}{\sqrt{2\omega_{k}}}e^{i\omega_{k}t}\dfrac{2ik}{2ik-\alpha}e^{ikx}a_{k}^{\dag}\bigg)\Bigg) \times \\
    \times\Bigg(\int\limits_{-\infty}^{0}\dfrac{dq}{2\pi}\bigg(\dfrac{-i\omega_{q}}{\sqrt{2\omega_{q}}}e^{-i\omega_{q}t}\big(e^{-iqx}+\dfrac{\alpha}{2iq-\alpha}e^{iqx}\big)a_{q}+\dfrac{i\omega_{q}}{\sqrt{2\omega_{q}}}e^{i\omega_{q}t}\big(e^{iqx}-\dfrac{\alpha}{2iq+\alpha}e^{-iqx}\big)a_{q}^{\dag}\bigg)+ \\
+\int\limits_{0}^{+\infty}\dfrac{dq}{2\pi}\bigg(\dfrac{-i\omega_{q}}{\sqrt{2\omega_{q}}}e^{-i\omega_{q}t}\dfrac{2iq}{2iq+\alpha}e^{-iqx}a_{q}+\dfrac{i\omega_{q}}{\sqrt{2\omega_{q}}}e^{i\omega_{q}t}\dfrac{2iq}{2iq-\alpha}e^{iqx}a_{q}^{\dag}\bigg)\Bigg)\Bigg].
 \end{multline}
Similarly:
\begin{multline}
\label{eq:s9}
   \int\limits_{-\infty}^{+\infty}dx(\partial_{x}\phi)^2=\int\limits_{-\infty}^{0}dx(\partial_{x}\phi)^2+\int\limits_{0}^{+\infty}dx(\partial_{x}\phi)^2= \\
   =\int\limits_{-\infty}^{0}dx\Bigg[\Bigg(\int\limits_{-\infty}^{0}\dfrac{dk}{2\pi}\bigg(\dfrac{-ik}{\sqrt{2\omega_{k}}}e^{-i\omega_{k}t}\dfrac{2ik}{2ik-\alpha}e^{-ikx}a_{k}+\dfrac{ik}{\sqrt{2\omega_{k}}}e^{i\omega_{k}t}\dfrac{2ik}{2ik+\alpha}e^{ikx}a^{\dag}_{k}\bigg)+ \\
   +\int\limits_{0}^{+\infty}\dfrac{dk}{2\pi}\bigg(\dfrac{-ik}{\sqrt{2\omega_{k}}}e^{-i\omega_{k}t}\big(e^{-ikx}+\dfrac{\alpha}{2ik+\alpha}e^{ikx}\big)a_{k}+\dfrac{ik}{\sqrt{2\omega_{k}}}e^{i\omega_{k}t}\big(e^{ikx}-\dfrac{\alpha}{2ik-\alpha}e^{-ikx}\big)a_{k}^{\dag}\bigg)\Bigg)\times \\
   \times\Bigg(\int\limits_{-\infty}^{0}\dfrac{dq}{2\pi}\bigg(\dfrac{-iq}{\sqrt{2\omega_{q}}}e^{-i\omega_{q}t}\dfrac{2iq}{2iq-\alpha}e^{-iqx}a_{q}+\dfrac{iq}{\sqrt{2w_{q}}}e^{i\omega_{q}t}\dfrac{2iq}{2iq+\alpha}e^{iqx}a^{\dag}_{q}\bigg)+ \\
   +\int\limits_{0}^{+\infty}\dfrac{dq}{2\pi}\bigg(\dfrac{-iq}{\sqrt{2\omega_{q}}}e^{-i\omega_{q}t}\big(e^{-iqx}+\dfrac{\alpha}{2iq+\alpha}e^{iqx}\big)a_{q}+\dfrac{iq}{\sqrt{2\omega_{q}}}e^{i\omega_{q}t}\big(e^{iqx}-\dfrac{\alpha}{2iq-\alpha}e^{-iqx}\big)a_{q}^{\dag}\bigg)\Bigg)\Bigg] + \\
   +\int\limits_{0}^{+\infty}dx\Bigg[\Bigg(\int\limits_{-\infty}^{0}\dfrac{dk}{2\pi}\bigg(\dfrac{-ik}{\sqrt{2\omega_{k}}}e^{-i\omega_{k}t}\big(e^{-ikx}-\dfrac{\alpha}{2ik-\alpha}e^{ikx}\big)a_{k}+\dfrac{ik}{\sqrt{2\omega_{k}}}e^{i\omega_{k}t}\big(e^{ikx}+\dfrac{\alpha}{2ik+\alpha}e^{-ikx}\big)a_{k}^{\dag}\bigg)+ \\
   +\int\limits_{0}^{+\infty}\dfrac{dk}{2\pi}\bigg(\dfrac{-ik}{\sqrt{2\omega_{k}}}e^{-i\omega_{k}t}\dfrac{2ik}{2ik+\alpha}e^{-ikx}a_{k}+\dfrac{ik}{\sqrt{2\omega_{k}}}e^{i\omega_{k}t}\dfrac{2ik}{2ik-\alpha}e^{ikx}a_{k}^{\dag}\bigg)\Bigg) \times \\
   \times\Bigg(\int\limits_{-\infty}^{0}\dfrac{dq}{2\pi}\bigg(\dfrac{-iq}{\sqrt{2\omega_{q}}}e^{-i\omega_{q}t}\big(e^{-iqx}-\dfrac{\alpha}{2iq-\alpha}e^{iqx}\big)a_{q}+\dfrac{iq}{\sqrt{2\omega_{q}}}e^{i\omega_{q}t}\big(e^{iqx}+\dfrac{\alpha}{2iq+\alpha}e^{-iqx}\big)a_{q}^{\dag}\bigg)+ \\
   +\int\limits_{0}^{+\infty}\dfrac{dq}{2\pi}\bigg(\dfrac{-iq}{\sqrt{2\omega_{q}}}e^{-i\omega_{q}t}\dfrac{2iq}{2iq+\alpha}e^{-iqx}a_{q}+\dfrac{iq}{\sqrt{2\omega_{q}}}e^{i\omega_{q}t}\dfrac{2iq}{2iq-\alpha}e^{iqx}a_{q}^{\dag}\bigg)\Bigg)\Bigg].
\end{multline}
And finally:
\begin{multline}
 \label{eq:s10}
       \int\limits_{-\infty}^{+\infty}m^{2}\phi^{2}dx=m^{2}\int\limits_{-\infty}^{0}dx(\phi)^2+m^{2}\int\limits_{0}^{+\infty}dx(\phi)^2= \\
       =m^{2}\int\limits_{-\infty}^{0}dx\Bigg[\Bigg(\int\limits_{-\infty}^{0}\dfrac{dk}{2\pi}\bigg(\dfrac{1}{\sqrt{2\omega_{k}}}e^{-i\omega_{k}t}\dfrac{2ik}{2ik-\alpha}e^{-ikx}a_{k}+\dfrac{1}{\sqrt{2\omega_{k}}}e^{i\omega_{k}t}\dfrac{2ik}{2ik+\alpha}e^{ikx}a^{\dag}_{k}\bigg)+ \\
       +\int\limits_{0}^{+\infty}\dfrac{dk}{2\pi}\bigg(\dfrac{1}{\sqrt{2\omega_{k}}}e^{-i\omega_{k}t}\big(e^{-ikx}-\dfrac{\alpha}{2ik+\alpha}e^{ikx}\big)a_{k}+\dfrac{1}{\sqrt{2\omega_{k}}}e^{i\omega_{k}t}\big(e^{ikx}+\dfrac{\alpha}{2ik-\alpha}e^{-ikx}\big)a_{k}^{\dag}\bigg)\Bigg)\times \\
       \times\Bigg(\int\limits_{-\infty}^{0}\dfrac{dq}{2\pi}\bigg(\dfrac{1}{\sqrt{2\omega_{q}}}e^{-i\omega_{q}t}\dfrac{2iq}{2iq-\alpha}e^{-iqx}a_{q}+\dfrac{1}{\sqrt{2\omega_{q}}}e^{i\omega_{q}t}\dfrac{2iq}{2iq+\alpha}e^{iqx}a^{\dag}_{q}\bigg)+ \\
       +\int_{0}^{+\infty}\dfrac{dq}{2\pi}\bigg(\dfrac{1}{\sqrt{2\omega_{q}}}e^{-i\omega_{q}t}\big(e^{-iqx}-\dfrac{\alpha}{2iq+\alpha}e^{iqx}\big)a_{q}+\dfrac{1}{\sqrt{2\omega_{q}}}e^{i\omega_{q}t}\big(e^{iqx}+\dfrac{\alpha}{2iq-\alpha}e^{-iqx}\big)a_{q}^{\dag}\bigg)\Bigg)\Bigg] + \\
       +m^{2}\int\limits_{0}^{+\infty}dx\Bigg[\Bigg(\int\limits_{-\infty}^{0}\dfrac{dk}{2\pi}\bigg(\dfrac{1}{\sqrt{2\omega_{k}}}e^{-i\omega_{k}t}\big(e^{-ikx}+\dfrac{\alpha}{2ik-\alpha}e^{ikx}\big)a_{k}+\dfrac{1}{\sqrt{2\omega_{k}}}e^{i\omega_{k}t}\big(e^{ikx}-\dfrac{\alpha}{2ik+\alpha}e^{-ikx}\big)a_{k}^{\dag}\bigg)+ \\
       +\int\limits_{0}^{+\infty}\dfrac{dk}{2\pi}\bigg(\dfrac{1}{\sqrt{2\omega_{k}}}e^{-i\omega_{k}t}\dfrac{2ik}{2ik+\alpha}e^{-ikx}a_{k}+\dfrac{1}{\sqrt{2\omega_{k}}}e^{i\omega_{k}t}\dfrac{2ik}{2ik-\alpha}e^{ikx}a_{k}^{\dag}\bigg)\Bigg) \times \\
       \times\Bigg(\int\limits_{-\infty}^{0}\dfrac{dq}{2\pi}\bigg(\dfrac{1}{\sqrt{2\omega_{q}}}e^{-i\omega_{q}t}\big(e^{-iqx}+\dfrac{\alpha}{2iq-\alpha}e^{iqx}\big)a_{q}+\dfrac{1}{\sqrt{2\omega_{q}}}e^{i\omega_{q}t}\big(e^{iqx}-\dfrac{\alpha}{2iq+\alpha}e^{-iqx}\big)a_{q}^{\dag}\bigg)+ \\
       +\int\limits_{0}^{+\infty}\dfrac{dq}{2\pi}\bigg(\dfrac{1}{\sqrt{2\omega_{q}}}e^{-i\omega_{q}t}\dfrac{2iq}{2iq+\alpha}e^{-iqx}a_{q}+\dfrac{1}{\sqrt{2\omega_{q}}}e^{i\omega_{q}t}\dfrac{2iq}{2iq-\alpha}e^{iqx}a_{q}^{\dag}\bigg)\Bigg)\Bigg].
\end{multline}

\subsubsection{The part of the Hamiltonian which contains $a_{k}a_{q}$}

Adding the terms containing $a_{k}a_{q}$ in $\eqref{eq:s8}$, $\eqref{eq:s9}$ and $\eqref{eq:s10}$ we obtain:
$$\Delta H_{aa}=\int\limits_{0}^{+\infty}dx\int\limits_{-\infty}^{0}\dfrac{dk}{2\pi}\int\limits_{-\infty}^{0}\dfrac{dq}{2\pi}a_{k}a_{q}\dfrac{e^{-it(\omega_{k}+\omega_{q})}}{\sqrt{4\omega_{k}\omega_{q}}}\Bigg[\big(\omega_{k}\omega_{q}+kq-m^{2}\big)\big(-e^{ix(k+q)}-e^{-ix(k+q)}\big)-$$
$$-\big(\omega_{k}\omega_{q}+kq-m^{2}\big)\dfrac{2i\alpha(k+q)}{(2ik-\alpha)(2iq-\alpha)}e^{ix(k+q)}-\alpha \big(\omega_{k}\omega_{q}-kq-m^{2}\big)\big[\dfrac{e^{-ix(k-q)}}{2iq-\alpha}+\dfrac{e^{ix(k-q)}}{2ik-\alpha}\big]\Bigg]-$$
$$-\int\limits_{0}^{+\infty}dx\int\limits_{-\infty}^{0}\dfrac{dk}{2\pi}\int\limits_{0}^{+\infty}\dfrac{dq}{2\pi}2a_{k}a_{q}\dfrac{e^{-it(\omega_{k}+\omega_{q})}}{\sqrt{4\omega_{k}\omega_{q}}}\times$$
$$\times\Bigg[\big(\omega_{k}\omega_{q}+kq-m^{2}\big)\bigg(\dfrac{2ik}{2ik-\alpha}e^{ix(k+q)}+\dfrac{2iq}{2iq+\alpha}e^{-ix(l+q)}\bigg)-\dfrac{2i\alpha (k-q)}{(2ik-\alpha)(2iq+\alpha)}\big(\omega_{k}\omega_{q}-kq-m^{2}\big)\Bigg]+$$

\begin{multline}
    +\int\limits_{0}^{+\infty}dx\int\limits_{0}^{+\infty}\dfrac{dk}{2\pi}\int\limits_{0}^{+\infty}\dfrac{dq}{2\pi}a_{k}a_{q}\dfrac{e^{-it(\omega_{k}+\omega_{q})}}{\sqrt{4\omega_{k}\omega_{q}}}\Bigg[\big(\omega_{k}\omega_{q}+kq-m^{2}\big)\big(-e^{-ix(k+q)}-e^{ix(k+q)}\big)+\\
    + \big(\omega_{k}\omega_{q}+kq-m^{2}\big)\dfrac{2i\alpha(k+q)}{(2ik+\alpha)(2iq+\alpha)}e^{-ix(k+q)}+\alpha \big(\omega_{k}\omega_{q}-kq-m^{2}\big)\big[\dfrac{e^{ix(k-q)}}{2iq+\alpha}+\dfrac{e^{-ix(k-q)}}{2ik+\alpha}\big]\Bigg]+\alpha \phi^{2}(t,0),
\end{multline}
where $[..]_{aa}$ and $\Delta H_{aa}$ means that we take only that part of the operator, which is proportional to $a_{k}a_{q}$.
Combining all contributions into one integral, we find:
\begin{multline}
\label{eq:008}
    \Delta H_{aa}=\int\limits_{0}^{+\infty}dx\int\limits_{0}^{+\infty}\dfrac{dk}{2\pi}\int\limits_{0}^{+\infty}\dfrac{dq}{2\pi}\dfrac{e^{-it(\omega_{k}+\omega_{q})}}{\sqrt{4\omega_{k}\omega_{q}}}\times \\
    \times \Bigg[ a_{-k}a_{-q}\big(\omega_{k}\omega_{q}+kq-m^{2}\big)\dfrac{2i\alpha(k+q)}{(2ik+\alpha)(2iq+\alpha)}e^{-ix(k+q)}+a_{-k}a_{-q}\alpha\big(\omega_{k}\omega_{q}-kq-m^{2}\big)\big[\dfrac{e^{ix(k-q)}}{2iq+\alpha}+\dfrac{e^{-ix(k-q)}}{2ik+\alpha}\big]- \\
    +2a_{-k}a_{q}\big(\omega_{k}\omega_{q}+kq-m^{2}\big)\dfrac{2i\alpha(k+q)}{(2ik+\alpha)(2iq+\alpha)}e^{-ix(k+q)}+2a_{-k}a_{q}\alpha\big(\omega_{k}\omega_{q}-kq-m^{2}\big)\big[\dfrac{e^{ix(k-q)}}{2iq+\alpha}+\dfrac{e^{-ix(k-q)}}{2ik+\alpha}\big]+ \\
    +a_{k}a_{q}\big(\omega_{k}\omega_{q}+kq-m^{2}\big)\dfrac{2i\alpha(k+q)}{(2ik+\alpha)(2iq+\alpha)}e^{-ix(k+q)}+\alpha a_{k}a_{q}\big(\omega_{k}\omega_{q}-kq-m^{2}\big)\big[\dfrac{e^{ix(k-q)}}{2iq+\alpha}+\dfrac{e^{-ix(k-q)}}{2ik+\alpha}\big]\Bigg]+\alpha \phi^{2}(t,0)= \\
    =\int\limits_{0}^{+\infty}dx\int\limits_{0}^{+\infty}\dfrac{dk}{2\pi}\int\limits_{0}^{+\infty}\dfrac{dq}{2\pi}\dfrac{e^{-it(w_{k}+w_{q})}}{\sqrt{4\omega_{k}\omega_{q}}}\times \\
    \times\Bigg[\bigg(\dfrac{2i\alpha(k+q)}{(2ik+\alpha)(2iq+\alpha)}e^{-ix(k+q)}\big(\omega_{k}\omega_{q}+kq-m^{2}\big)+\alpha\bigg(\dfrac{e^{ix(k-q)}}{2iq+\alpha}+\dfrac{e^{-ix(k-q)}}{2ik+\alpha}\bigg)\big(\omega_{k}\omega_{q}-kq-m^{2}\big)\bigg)\times \\
    \times \big(a_{-k}a_{-q}+2a_{-k}a_{q}+a_{k}a_{q}\big)\Bigg]+\alpha \phi^{2}(t,0).
\end{multline}
Integrating over the $x$ the first term in the last line of \eqref{eq:008}, we get:
\begin{multline}
\label{eq:009}
    \dfrac{2i\alpha(k+q)}{(2ik+\alpha)(2iq+\alpha)}\big(\omega_{k}\omega_{q}+kq-m^{2}\big)\int\limits_{0}^{+\infty}dxe^{-ix(k+q)}= \\
    =\dfrac{2i\alpha(k+q)}{(2ik+\alpha)(2iq+\alpha)}\dfrac{1}{i(k+q-i\epsilon)}\big(\omega_{k}\omega_{q}+kq-m^{2}\big)=\dfrac{2\alpha(k+q)}{(2ik+\alpha)(2iq+\alpha)}\dfrac{(k+q)+i\epsilon}{(k+q)^2+\epsilon^{2}}\big(\omega_{k}\omega_{q}+kq-m^{2}\big)= \\
    =\dfrac{2\alpha}{(2ik+\alpha)(2iq+\alpha)}\big(\omega_{k}\omega_{q}+kq-m^{2}\big)+\dfrac{2\alpha(k+q)}{(2ik+\alpha)(2iq+\alpha)}i\pi\delta(k+q)\big(\omega_{k}\omega_{q}+kq-m^{2}\big)= \\
    =\dfrac{2\alpha}{(2ik+\alpha)(2iq+\alpha)}\big(\omega_{k}\omega_{q}+kq-m^{2}\big).
\end{multline}
Integrating over the $x$ the second term in the last line of \eqref{eq:008}, we get:
\begin{multline}
\label{eq:010}
    \big(\omega_{k}\omega_{q}-kq-m^{2}\big)\int\limits_{0}^{+\infty}\bigg(\dfrac{e^{ix(k-q)}}{2iq+\alpha}+\dfrac{e^{-ix(k-q)}}{2ik+\alpha}\bigg)= \\
    =\big(\omega_{k}\omega_{q}-kq-m^{2}\big)\bigg(\dfrac{-1}{i(2iq+\alpha)(k-q+i\epsilon)}+\dfrac{1}{i(2ik+\alpha)(k-q-i\epsilon)}\bigg)= \\
    =\dfrac{i\big(\omega_{k}\omega_{q}-kq-m^{2}\big)}{(2iq+\alpha)(2ik+\alpha)}\dfrac{(2i(k-q)^2+2\epsilon(k-q)-2i\alpha \epsilon)}{(k-q)^2+\epsilon^{2}}=-2\dfrac{\big(\omega_{k}\omega_{q}-kq-m^{2}\big)}{(2iq+\alpha)(2ik+\alpha)}.
\end{multline}
Also:
\begin{equation}
  \label{eq:007}
   \Big[\alpha \phi^{2}(t,0)\Big]_{aa}=-\int\limits_{0}^{+\infty}\dfrac{dk}{2\pi}\int\limits_{0}^{+\infty}\dfrac{dq}{2\pi}\dfrac{e^{-it(w_{k}+w_{q})}}{\sqrt{\omega_{k}\omega_{q}}}\dfrac{4\alpha kq}{(2iq+\alpha)(2ik+\alpha)}\big(a_{-k}a_{-q}+2a_{-k}a_{q}+a_{k}a_{q}\big).
\end{equation}
Then using \eqref{eq:009}-\eqref{eq:007}, finally we find that:
\begin{multline}
   \Delta H_{aa}=\int\limits_{0}^{+\infty}\dfrac{dk}{2\pi}\int\limits_{0}^{+\infty}\dfrac{dq}{2\pi}\dfrac{e^{-it(\omega_{k}+\omega_{q})}}{\sqrt{4\omega_{k}\omega_{q}}}\times \\
    \times\Bigg[\dfrac{2\alpha}{(2iq+\alpha)(2ik+\alpha)}\bigg(\omega_{k}\omega_{q}+kq-m^{2}-\big(\omega_{k}\omega_{q}-kq-m^{2}\big)\bigg) \big(a_{-k}a_{-q}+2a_{-k}a_{q}+a_{k}a_{q}\big)\Bigg]+\alpha \phi^{2}(t,0)= \\
    =\int\limits_{0}^{+\infty}\dfrac{dk}{2\pi}\int\limits_{0}^{+\infty}\dfrac{dq}{2\pi}\dfrac{e^{-it(\omega_{k}+\omega_{q})}}{\sqrt{4\omega_{k}\omega_{q}}}\dfrac{4\alpha kq}{(2iq+\alpha)(2ik+\alpha)}\big(a_{-k}a_{-q}+2a_{-k}a_{q}+a_{k}a_{q}\big)- \\
    -\int\limits_{0}^{+\infty}\dfrac{dk}{2\pi}\int\limits_{0}^{+\infty}\dfrac{dq}{2\pi}\dfrac{e^{-it(\omega_{k}+\omega_{q})}}{\sqrt{4\omega_{k}\omega_{q}}}\dfrac{4\alpha kq}{(2iq+\alpha)(2ik+\alpha)}\big(a_{-k}a_{-q}+2a_{-k}a_{q}+a_{k}a_{q}\big)=0.
\end{multline}
Thus, in the case under consideration the Hamiltonian does not contain neither $aa$ nor $a^{\dagger}a^{\dagger}$ terms.

\subsubsection{The part of the Hamiltonian which contains $a_{k}a^{\dag}_{q}$}

Similarly to the above derivation, adding $a_{k}a^{\dagger}_{q}$-terms from $\eqref{eq:s8}$, $\eqref{eq:s9}$, $\eqref{eq:s10}$ and
\begin{equation}
    \label{eq:25}
    \Big[\alpha \phi^{2}(t,0)\Big]_{aa^{\dag}}=
\end{equation}
$$=-\int\limits_{-\infty}^{0}\dfrac{dk}{2\pi}\int\limits_{-\infty}^{0}\dfrac{dq}{2\pi}a_{k}a^{\dag}_{q}\dfrac{e^{-it(\omega_{k}-\omega_{q})}}{\sqrt{4\omega_{k}\omega_{q}}}\dfrac{4\alpha kq}{(2ik-\alpha)(2iq+\alpha)}-$$
$$-\int\limits_{-\infty}^{0}\dfrac{dk}{2\pi}\int\limits_{0}^{+\infty}\dfrac{dq}{2\pi}\dfrac{1}{\sqrt{4\omega_{k}\omega_{q}}}\Bigg[\dfrac{4\alpha kq e^{-it(\omega_{k}-\omega_{q})}}{(2ik-\alpha)(2iq-\alpha)}a_{k}a^{\dag}_{q}+\dfrac{4\alpha kq e^{-it(\omega_{q}-\omega_{k})}}{(2ik+\alpha)(2iq+\alpha)}a_{q}a^{\dag}_{k}\Bigg]-$$
$$-\int\limits_{0}^{+\infty}\dfrac{dk}{2\pi}\int\limits_{0}^{+\infty}\dfrac{dq}{2\pi}a_{k}a^{\dag}_{q}\dfrac{e^{-it(\omega_{k}-\omega_{q})}}{\sqrt{4\omega_{k}\omega_{q}}}\dfrac{4\alpha kq}{(2ik+\alpha)(2iq-\alpha)},$$
we get that:
\begin{equation}
    \label{eq:26}
    H_{aa^{\dag}}=\int\limits_{-\infty}^{+\infty} \dfrac{dk}{2\pi} \frac{\omega}{2} a_{k}a^{\dag}_{k}.
\end{equation}
Thus, the free Hamiltonian is:
\begin{equation}
    \label{eq:27}
    H=\int\limits_{-\infty}^{+\infty} \dfrac{dk}{2\pi} \frac{\omega}{2} \big(a_{k}a^{\dag}_{k}+a^{\dag}_{k}a_{k}\big),
\end{equation}
i.e. is diagonal as is expected.

\subsection{The vacuum expectation value of the stress-energy tensor}

Using the same regularization as for the ideal mirror, we find the following expressions for the components of the vacuum expectation value of the stress-energy tensor:
$$\braket{T_{tx}}=\int\limits_{-\infty}^{+\infty}dkk\frac{k^2}{k^2+\alpha^2}=0,$$
$$\braket{T_{tt}}=-\frac{1}{4\pi}M^2\log{\Lambda}+\frac{\alpha\delta(x)}{2\pi}\bigg[\frac{\arctan(\sqrt{\frac{4M^2}{\alpha^2}-1})}{\sqrt{\frac{4M^2}{\alpha^2}-1}}
-\frac{\arctan(\sqrt{\frac{4m^2}{\alpha^2}-1})}{\sqrt{\frac{4m^2}{\alpha^2}-1}}\bigg],$$
$$\braket{T_{xx}}=\frac{1}{4\pi}M^2\log{\Lambda}-\frac{\alpha\delta(x)}{2\pi}\bigg[\frac{\arctan(\sqrt{\frac{4M^2}{\alpha^2}-1})}{\sqrt{\frac{4M^2}{\alpha^2}-1}}
-\frac{\arctan(\sqrt{\frac{4m^2}{\alpha^2}-1})}{\sqrt{\frac{4m^2}{\alpha^2}-1}}\bigg]+F(m,M,x),$$
where $$F(m,M,x)=2m^2\int\limits_{-\infty}^{+\infty}\bigg[\theta(-x)\theta(k)\Big(\frac{\alpha}{2ik-\alpha}e^{-2ikx}-\frac{\alpha}{2ik+\alpha}e^{2ikx}\Big)+$$$$
+\theta(x)\theta(-k)\Big(\frac{\alpha}{2ik-\alpha}e^{2ikx}-\frac{\alpha}{2ik+\alpha}e^{-2ikx}\Big)\bigg]\frac{dk}{4\pi\omega}-(m\rightarrow M).$$
In the limit $M\to\infty$ we have that $\frac{\arctan(\sqrt{\frac{4M^2}{\alpha^2}-1})}{\sqrt{\frac{4M^2}{\alpha^2}-1}}\to0$. Hence,
\begin{equation}
\braket{T_{\mu\nu}}=-\frac{1}{4\pi}\eta_{\mu\nu}\bigg[M^2\log{\Lambda}+2\alpha\delta(x)
\frac{\arctan(\sqrt{\frac{4m^2}{\alpha^2}-1})}{\sqrt{\frac{4m^2}{\alpha^2}-1}}\bigg]+F(m,M,x)\begin{pmatrix} 0&0\\
0&1\end{pmatrix}.
\end{equation}
One can see that $F(m,M,x)\to0$, as $x\to\infty$. To estimate $F(m,M,x)$, as $x\to0$, when $M\gg m$, and $2m\neq \alpha$:
$$\lim_{x\to0}F(m,M,x)=\frac{M^2}{\pi}\frac{\arctan(\sqrt{\frac{4M^2}{\alpha^2}-1})}{\sqrt{\frac{4M^2}{\alpha^2}-1}}-
\frac{m^2}{\pi}\frac{\arctan(\sqrt{\frac{4m^2}{\alpha^2}-1})}{\sqrt{\frac{4m^2}{\alpha^2}-1}}\to\alpha\frac{M}{4},$$
and for the case, when $2m=\alpha$, we obtain that:
$$\lim_{x\to0}F(m,M,x)=\frac{M^2}{\pi}\frac{\arctan(\sqrt{\frac{4M^2}{\alpha^2}-1})}{\sqrt{\frac{4M^2}{\alpha^2}-1}}-\frac{m^2}{\pi}\to m\frac{M}{2}$$
For other values of $x$ $F(m,M,x)$ is finite. After the boost we find the following vacuum expectation value:
$$\braket{T_{t'x'}}=\beta\gamma^2F(m,M,x'+\beta t')$$ which is infinite on the mirror, as $x'\to-\beta t'$, and tends to zero far away from the delta-potential world-line. Similarly to the case of ideal mirror, we observe that the $\delta$-functional potential somehow captures and carries a portion of the zero-point fluctuations along with itself.

\section{Conclusion}

The main results of the present paper are as follows.

First, the ideal mirror, which reflects all the modes equally well, is rather pathological situation from the physical point of view. That is at least because of the fact that field operator and its conjugate momentum do not have canonical commutation relations. That is true under assumption that the creation and annihilation operators obey the standard Heisenberg algebra. The problems become even stronger in the case of loops in the interacting field theory \cite{Akhmedov:2018lkp}.

Second, in the presence of moving mirrors the diagonal form (in terms of the creation and annihilation operators) has the $H-\beta P$ operator rather than $H$ itself. Here $\beta$ is the velocity of the mirror and $P$ is the momentum operator.

Third, for the massive fields in the presence of a mirror moving with constant velocity the expectation value of the stress--energy tensor has a non--diagonal contribution. This is not a flux, because it decays with the increase of the distance from the mirror. Such a contribution is present both in the case of ideal and non--ideal mirror. It appears due to the fact that moving mirror distorts somehow zero point fluctuations of massive fields and the violation of the Lorentz invariance in the presence of a mirror becomes apparent. Note that, on the contrary, massless fields always reside on the light--cone (both falling and reflecting waves) which is not affected by the presence of a reflecting boundary.

Fourth, in the case of non--ideal mirror the commutation relations of the field operator and its conjugate momentum have their canonical form, as it should be in proper physical situations.

We would like to thank Sergey Alexeev for the useful discussions and Emil T. Akhmedov for formulating this problem for us and for the sharing of his ideas. The work of Lev Astrakhantsev is performed under the financial support by the Russian state grant Goszadanie 3.9904.2017/8.9. Also this work has been funded by the Russian Academic Excellence Project '5-100'.

\end{document}